\documentclass[a4paper,11pt]{article}
\usepackage{jinstpub} % for details on the use of the package, please
\pdfoutput=1 % if your are submitting a pdflatex (i.e. if you have
\usepackage{lineno}
\modulolinenumbers[5]

\usepackage{xcolor}
\usepackage{graphicx}
\graphicspath{{pics_paper/}}

\usepackage{caption}
\usepackage{subcaption}

\usepackage{hyperref}
\usepackage{chemformula}

\title{Characterization of the BOLDPET optical prototype, an innovative Cherenkov detector for 511~keV $\gamma$ radiation.}

\author[a]{R.~Chyzh}
\author[a]{G.~Tauzin}
\author[a,b]{D.~Yvon}
\author[a,1]{, C.-H. Sung\note{Currently at GE Healthcare}}
\author[c]{D.~Breton} 
\author[c]{J.~Maalmi}
\author[d]{K.~Sch\"afers}
\author[e]{C.~Weinheimer}
\author[a,b,2]{and V.~Sharyy\note{Corresponding author}}

\affiliation[a]{IRFU, CEA,  Universit\'e Paris-Saclay,  Gif-sur-Yvette, France}
\affiliation[b]{BioMAPs, Service Hospitalier Fr\'ed\'eric Joliot, CEA, CNRS, Inserm, Universit\'e Paris-Saclay, Orsay, France}
\affiliation[c]{IJCLab, IN2P3, CNRS, Universit\'e Paris-Saclay, Orsay, France}
\affiliation[d]{European Institute for Molecular Imaging, University of M\"unster, M\"unster, Germany}
\affiliation[e]{Nuclear Physics Institute, University of M\"unster, M\"unster, Germany}

\emailAdd{viatcheslav.sharyy@cea.fr}

\abstract{
In the present work  we describe the design, construction, and testing of the optical prototype developed for the BOLDPET project,
with the objective of creating a PET detection module with high spatial and time resolution. 
The BOLDPET technology uses an innovative detection liquid, trimethylbismuth, for detecting 
511~keV $\gamma$-quanta resulting from positron annihilation.  
The optical signal is exclusively produced  through the Cherenkov mechanism, and the produced photons are detected using
Planacon  microchannel-plate photomultiplier.
We achieve an excellent time resolution of  150~ps (FWHM) within a sizable detection volume 
measuring 55 mm x 55 mm x 25 mm.  
Through detailed Geant4 simulations, we examine the limiting factors affecting time resolution and explore potential avenues for improvement.
Furthermore, we demonstrate the feasibility of coarse 2D localization of interactions using the optical signal alone, achieving a precision of about  5-8 mm (FWHM) within the homogeneous detection volume.
}


\begin{document}

\maketitle

% section 1
%\linenumbers

\section {Introduction} 
Positron Emission Tomography (PET) is a non-invasive medical imaging
technique  allowing to visualize  a radioactive tracer in a patient
body with a sensitivity down to picomole level.  It is 
widely used in oncology, cardiology, neurology, and biomedical research~\cite{Vallabhajosula_2011,Martinez2019Jun,Djekidel2022Sep}.
Despite its sensitivity, PET has a modest spatial resolution of 3-4 mm (FWHM\footnote{FWHM: Full Width at Half Maximum}) for whole-body or brain-sized scanners~\cite{HRRT_deJong_2007,Gonzalez2018Jun,Vandenberghe2020Dec,Zeimpekis2022Jul}. 
However, the physical limits imposed by the positron range and two gamma acollinearity suggest that PET can achieve a resolution of less than 1 mm~\cite{Saanchez-Crespo2004,Harpen2004Jan,Moses2011a,Lehnert2011May,Emond2019Oct}. 
This creates an opportunity for the development of more accurate scanners,
which are of significant interest, particularly for brain studies.  

One approach to achieving high spatial resolution is to use a matrix of scintillating crystals with a typical size of 1 mm x 1 mm and read them out individually. It is realized in the  preclinical-size PET scanners reaching the spatial resolution of about 1~mm,
see e.g. \cite{Belcari2017Jun, Amirrashedi2019Nov, Gaudin2021Mar}. 
However, this method requires a high number of electronics channels and is expensive when applied to a brain-sized scanner. 
Another challenge arises in the requirement for utilizing small voxels  (i.e., 1 mm$^3$) for image reconstruction.
This demands either augmenting statistics through prolonged scan times or elevated injected activity, or refining the signal-to-noise ratio by implementing the time-of-flight (TOF) technique~\citep{Vandenberghe_2016,Lecoq2020Oct,Schaart2021Apr}. 
Moreover, the ability to  measure  depth-of-interaction  coordinate in the detection module decreases the parallax
error during reconstruction and improves spatial resolution~\cite{Mohammadi2019}.
In recent years, the use of monolithic crystals  has gained significant interest as a promising solution to address 
these challenges in high-precision PET applications~\cite{Borghi_2016b,Gonzalez-Montoro2017,Krishnamoorthy2018Jul,Yvon2020Jul,Stockhoff2021Jul}.
In this approach, the entire profile of light emission on the crystal surface is detected,
enabling the reconstruction of 3D spatial coordinates with millimeter-scale precision,
along with a time measurement for TOF calculation.

\begin{figure}
    \centering\includegraphics[width=.7\columnwidth]{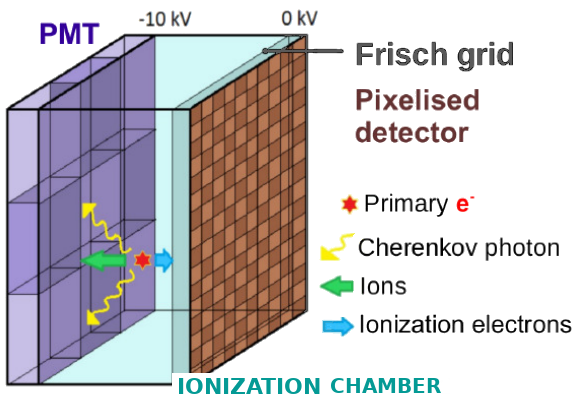}
    \caption{\label{fig:BOLDPET} Principle scheme of the BOLDPET detection principals~\cite{Ramos_2016}.
      The primary electron created by 511~keV $\gamma$ produces Cherenkov light and ionizes the medium.}
\end{figure}

We study an alternative approach in the BOLDPET project  (Bismuth Organometallic Liquid Detector for Positron Emission Tomography),
development of the previously proposed CaLIPSO concept \cite{Yvon2014}.
We propose to use a heavy liquid media, trimethylbismuth (TMBi), for
detecting the $\gamma$-quanta from the positron annihilation. TMBi liquid has a short, 25 mm, attenuation length and high, 47\%,
\cite{Hubbell2010XCOMP},
photoelectric fraction for converting the 511~keV photon into a relativistic electron.
This  electron produces about 20  Cherenkov photons  and ionizes the medium.
The detector operates as a time-projection chamber and detects both light and charge
signals, see Fig.~\ref{fig:BOLDPET} and \cite{Ramos2015}.

Ionization signal is read-out in  a classical Frisch grid configuration using densely pixelated anode matrix
and is used to determine precisely the 2D position of the $\gamma$ interaction. 
The optical photons provide a precise timing to determine the 3D coordinate using the ionization drift time and 
to improve the quality of the image reconstruction with the time-of-flight technique.
Such detector has the potential to reconstruct the 3D coordinate with a precision of 1~mm$^3$ or better and reach the time resolution of about 150~ps (FWHM).
Such characteristics give an interesting perspective of using this device as a brain-size PET scanner~\cite{Kochebina2018Nov} or in pre-clinical research.

Detection of the ionization signal is the most challenging part of this project
and requires an excellent purity of the liquid together with the special precautions to avoid the electrical break-down phenomena.
At the initial stage of the project, the detection of the ionization signal is studied using simplified test devices, and detailed discussions of the results can be found elsewhere~\cite{Ramos2015, Farradeche2019, Gerke2022Sep,Peters2022Sep}.
In this article, we exclusively focus on the investigation of the optical Cherenkov signal detection,
utilizing a dedicated optical prototype.

The first test of the optical detection~\cite{Ramos_2016} demonstrates good, 
35~\%, efficiency to detect 511~keV photons,  but achieved non-optimal time resolution
of 540~ps.
In this work we optimize the design of the optical prototype, in particular reduce the thickness of the TMBi sensitive volume  from 5~cm to 2.5~cm,
remove separation of the sensitive volume in cells and replace the conventional photo-multiplication tube (PMT) with a
micro-channel-plate (MCP) PMT, see section~\ref{sec:prototype} for details.
As a result, we were able to achieve a significant improvement in the time resolution (section~\ref{sec:time}), and demonstrate the ability to
have a coarse localization of the $\gamma$-conversion vertex (section~\ref{sec:spatial}).

%section 2
\section{Setup}
\label{sec:prototype}
The BOLDPET optical  prototype is a sealed reservoir for TMBi liquid that
serves as a Cherenkov radiator for relativistic  electrons produced by the conversion of 511~keV 
$\gamma$-quanta. TMBi is a transparent, chemically reactive, high density (2.3 g/cm$^3$) liquid
that was chosen to optimize photoelectric conversion, 
primarily due to the high atomic number, Z=83, of Bismuth. 
The refraction index of TMBi at a 450 nm wavelength was measured to be 1.62$\pm$0.06 in a previous study~\cite{Ramos2015}.

\begin{figure}[ht]
\begin{minipage}[t]{0.50\columnwidth}
\centering\includegraphics[width=\textwidth]{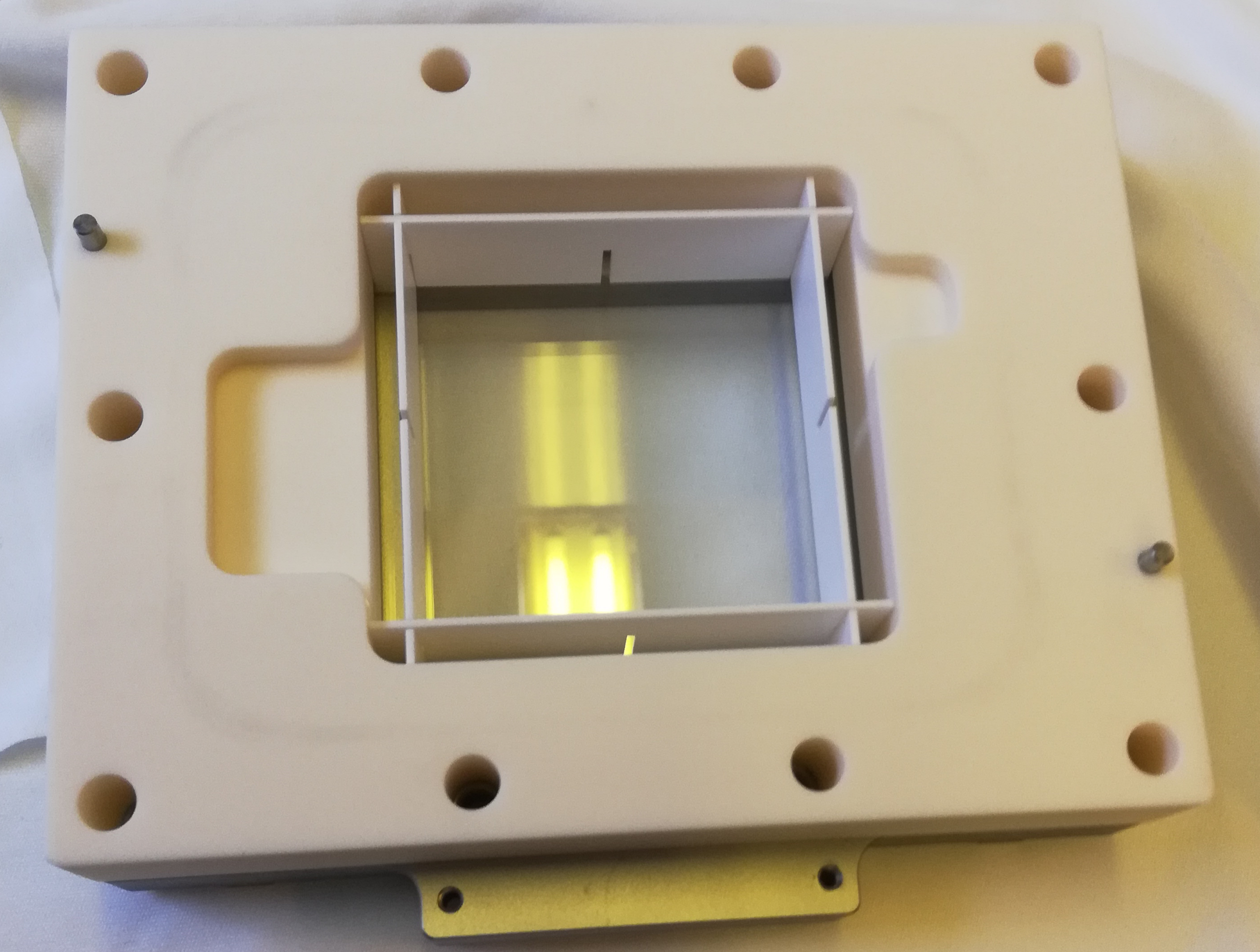}
\caption{\label{fig:Prot1}
The optical prototype at the assembling stage. Ceramic body and four diffusive white reflectors inside the body are visible.}
\end{minipage}
\hfill
\begin{minipage}[t]{0.47\columnwidth}
\centering \includegraphics[width=\textwidth]{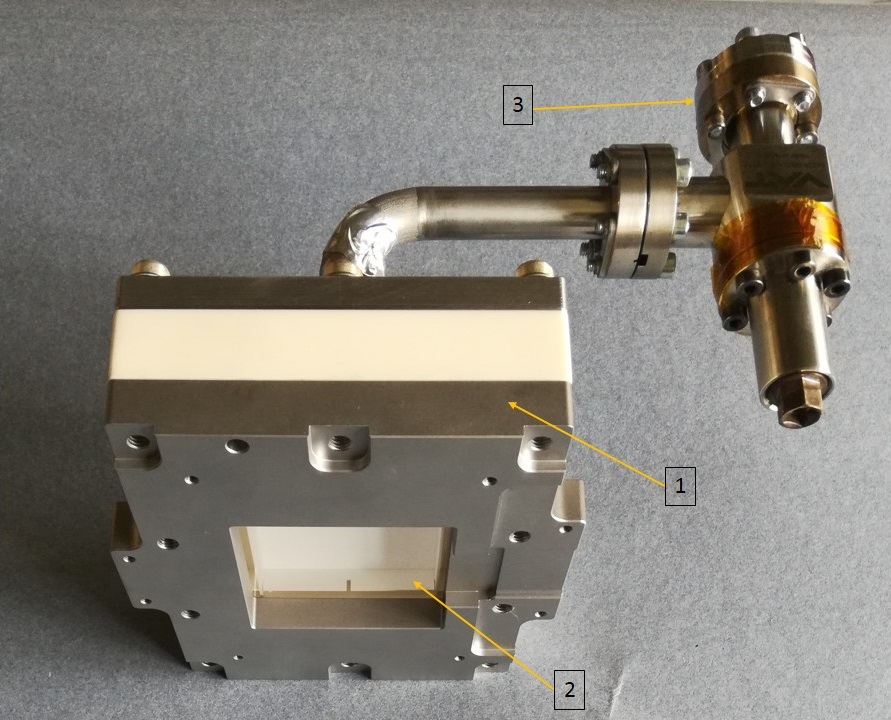}
\caption{\label{fig:Prot2}
The optical prototype filled with the TMBi liquid. Main elements: 1) Metal-ceramic body; 
2) Sapphire glass window; 3) Valve.}
\end{minipage}
\end{figure}

The high reactivity of TMBi necessitates that all manipulations and operations be conducted under ultra-high vacuum conditions. Any leak of air inside the prototype will induce a chemical reaction, 
leading to the transformation of the liquid TMBi into bismuth oxide powder. 
For this reason, for the prototype construction we have utilized materials known to be ultra-high vacuum compatible. 
Figures~\ref{fig:Prot1} and \ref{fig:Prot2} show the prototype,  with the body  made from 99.7\% alumina ceramic.
Hermeticity is achieved through the use of metallic gaskets compressed between the stainless steel cap and the ceramic body. To improve photon collection efficiency, we have integrated specifically chosen white diffusive ceramic reflectors within the liquid. 
The size of the detection volume was chosen to be 55~mm x 55~mm x 25~mm, matching the dimensions of the MCP-PMT window.
For the optical window, sapphire has been employed due to its high mechanical strength and a high refractive index, optimizing light transfer between the detection medium and the photomultiplier.

%subsection 2.2
\subsection{Photon Detection and Readout System}
\begin{figure}[ht]
\centering
 \includegraphics[width=.8\columnwidth]{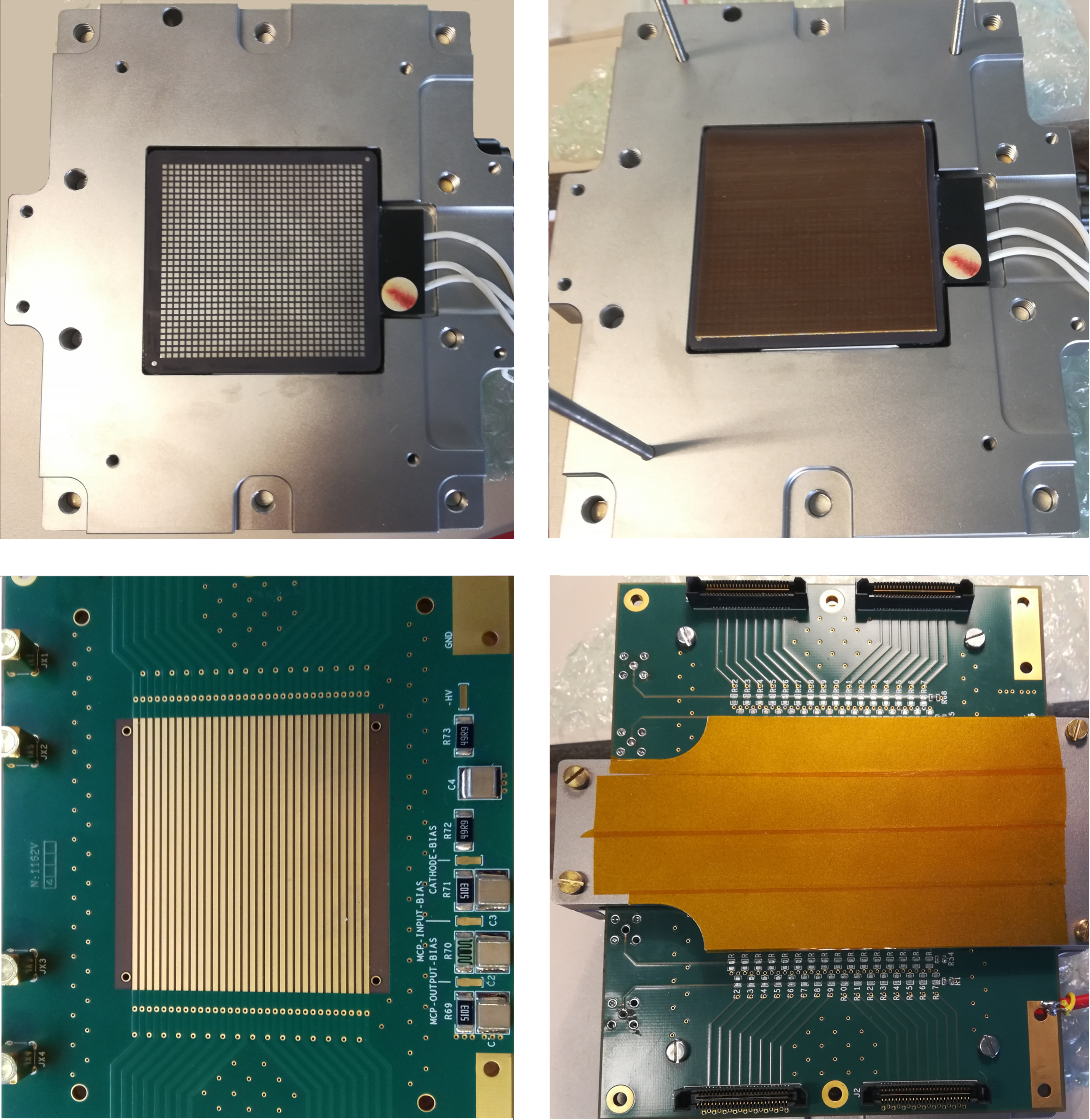}
 \caption{\label{fig:interconnect}
  Planacon MCP-PMT inserted into the optical prototype (top left) with the Inter-Connector interface positioned on top of the PMT (top right). 
  The transmission line PCB (bottom left). The transmission line PCB mounted on top of the PMT and pressurized using a metallic clamp (bottom right).
  }
\end{figure}

For the optical photons detection we have used Photonis's micro-channel plate photomultiplier tube (MCP-PMT)  Planacon XP85122.
%~\cite{XP85122}. 
To ensure optimal optical coupling we utilised Nye SmartGel's  gel OCF452H~\cite{SMARTGEL}.
This MCP-PMT has a pore diameter of 10~$\mu m$ and  an active photocathode area of  53x53~mm$^2$.
PMT's anode structure consists of  32x32 metallic pads of 1.1x1.1~mm$^2$ 
and a gap of 0.5~mm between the pads.
In order to limit the number of electronics channels
but keep high time resolution, we developed a read-out system using 50~Ohm 32 transmission lines 
printed on the 3-mm thick PCB. This PCB was connected to the anode pads with the pressure-sensitive anisotropic conductive sheet, Shin-Etsu Inter-Connector MT-4X  \cite{Interconnect}, see Fig.~\ref{fig:interconnect}.

\begin{figure}[ht]
\centering \includegraphics[width=.7\columnwidth]{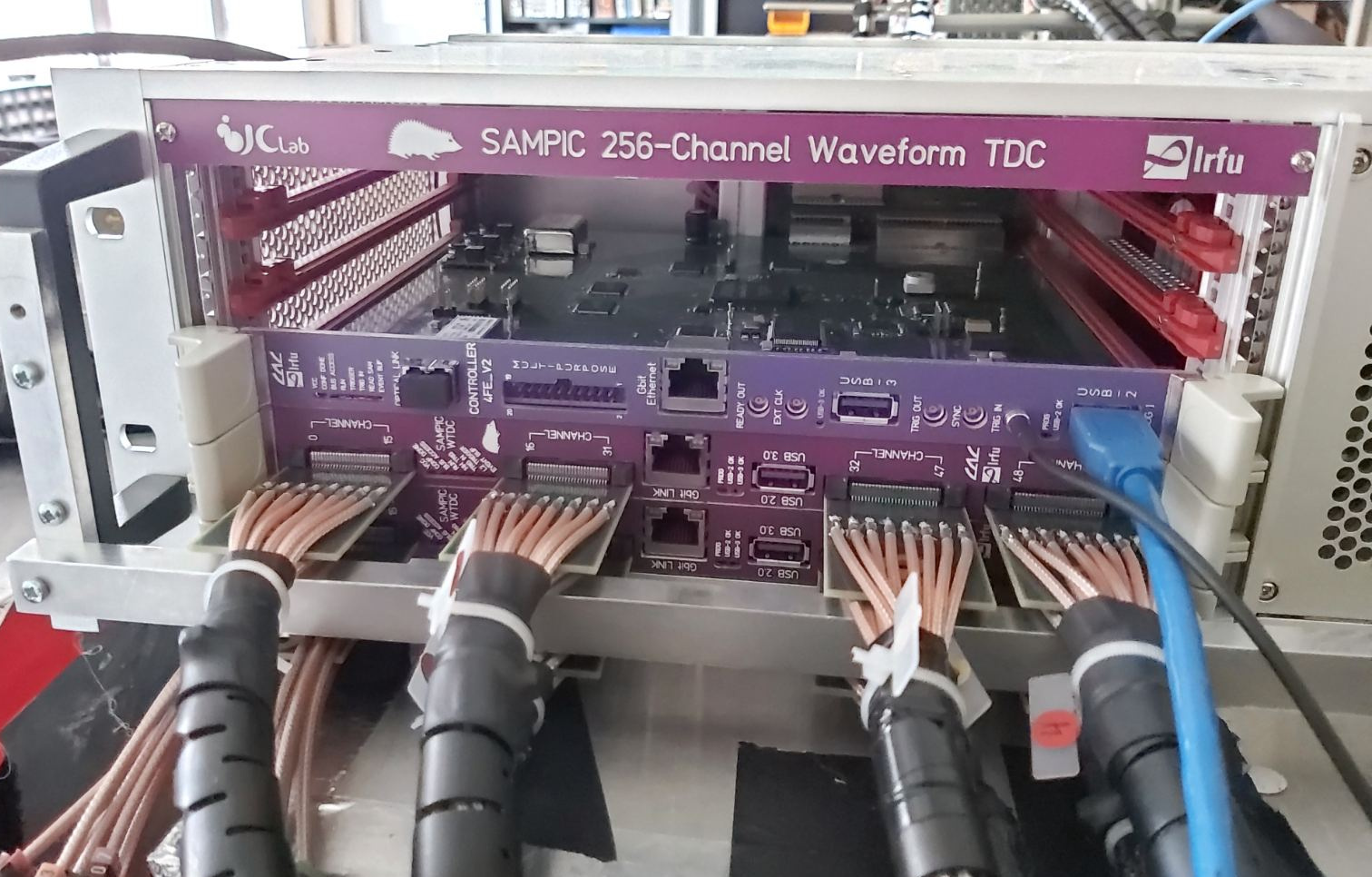}
\caption{\label{fig:sampic} The SAMPIC crate used for digitizing the data}
\end{figure}

We biased the PMT at a high voltage providing a gain of about $10^6$. 
At this gain, the single photoelectron  signal needs to be amplified by a factor of 100 to be in the range suitable for the digitization module. We used a two stage amplification using dedicated boards (700~MHz bandwidth, 2x20~dB amplification).  
For this test, we utilized a SAMPIC crate, Fig.~\ref{fig:sampic}, capable of accommodating up to four 64-channels SAMPIC modules. 
These modules use SAMPIC\_V3C chips, which are based on the patented concept of waveform and time-to-digital converter
\citep{Delagnes2014Nov,Delagnes:2015oda,Breton2020}.
The signal sampling frequency can be varied within a range of 1.6 to 8.5 GS/s (6.4 GS/s frequency is adopted in this work).
The digitization of individual events is performed over groups of 64 samples,
which are then used for accurate timing measurements on-line, but can also be stored for off-line processing.
SAMPIC board allows to reach a timing accuracy of approximately 5~ps (standard deviation)~\citep{Breton2016}.
More details about the read-out system can be found in~\cite{Follin2022Mar}. 

\subsection {The reference detector}
\begin{figure}[ht]
\begin{minipage}[t]{0.48\columnwidth}
\centering\includegraphics[width=\textwidth]{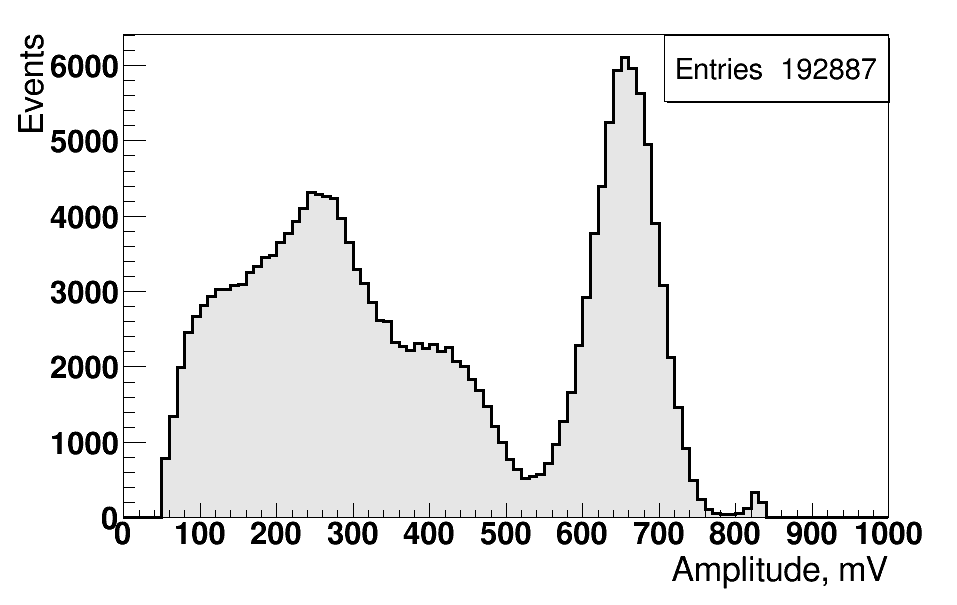}
\caption{\label{fig:sipm_amplitude}
  Amplitude spectrum in the reference LYSO+SiPM detector. The peak at 650 mV corresponds to the photo-ionization conversion of
  511~keV gamma rays.}
\end{minipage}
\hfill
\begin{minipage}[t]{0.48\columnwidth}
\centering\includegraphics[width=\textwidth]{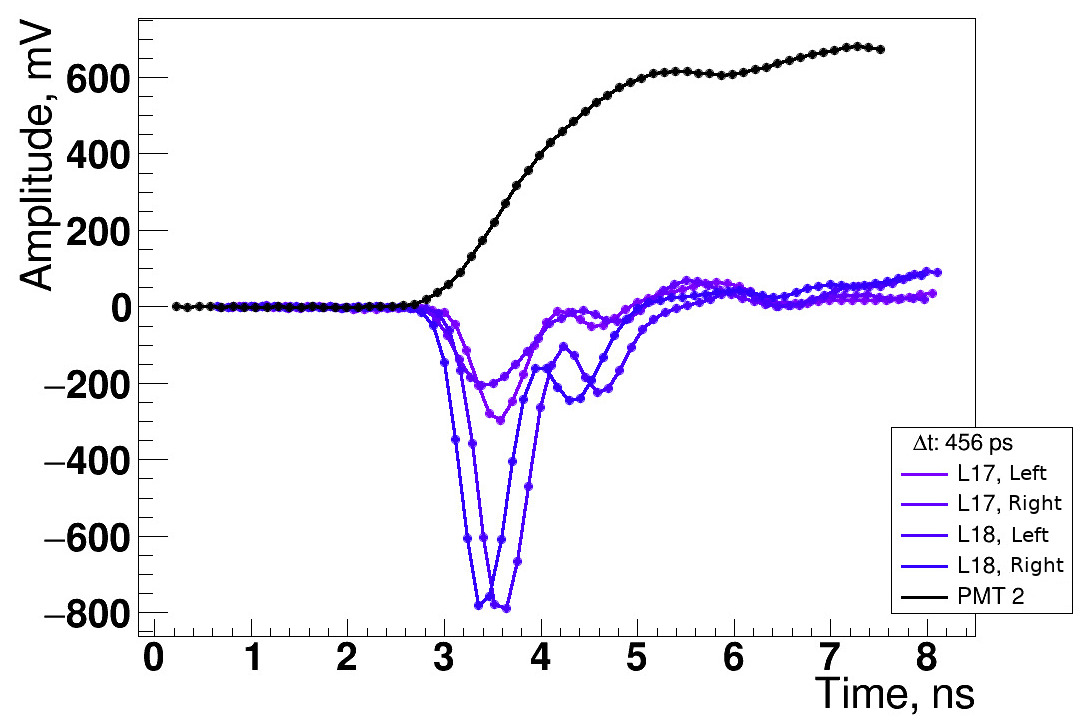}
\caption{\label{fig:coincidence}
  Typical signals in one event.
  Points represent the 64 samples digitized by the SAMPIC module.
  Signal with positive amplitude corresponds to the reference detector (PMT2).
  Other signals correspond to the left and right sides of two transmission lines of the prototype (L17, L18).
  $\Delta t$ is the time difference between the prototype and the reference detector.}
\end{minipage}
\end{figure}

The reference detector is assembled using a LYSO:Ca,Ce co-doped crystal provided by Saint-Gobain (France), with dimensions of 3$\times$3$\times$3 mm$^{3}$. This crystal is known for its improved rise time, excellent light yield, decay time, and afterglow characteristics \cite{LYSO}. 
The crystal is optically coupled using Histomount glue, diluted with 2/3 of Xylène to a 3$\times$3 mm$^2$ SENSL/ONSEMI silicon photo-multiplier (SiPM) MICROFC-30035-SMT-TR \cite{SENSL_SiPM}.
The assembly is wrapped in Teflon tape to increase the light collection efficiency. 
The SiPM is operating at $-30$~V bias (18\% overvoltage). The signal is amplified with a 20~dB, 2.7 GHz commercial amplifier and digitized with the SAMPIC module. Due to the limited number of available digitization samples, 64, 
we are able to record only the initial part of the detector signal. Despite this limitation,we achieve a 6.4\%  
resolution in amplitude (Standard Deviation)\footnote{This resolution is measured as the width of the histogram at half the maximum of the peak,
  divided by the peak amplitude, 650 mV. To convert it to standard deviation, we divide this value by 2.36.},
allowing clear differentiation between the photoinization peak and compton scattering events,
Fig.~\ref{fig:sipm_amplitude}. 
For determining the signal time, a fixed threshold time estimator is employed. The time resolution of this detector is measured to be 105~ps (FWHM).

\subsection{Tests with the radioactive source}
\label{sec:source}
\begin{figure}[ht]
  \begin{minipage}[ht]{0.6\columnwidth}
    \centering \includegraphics[width=0.7\textwidth]{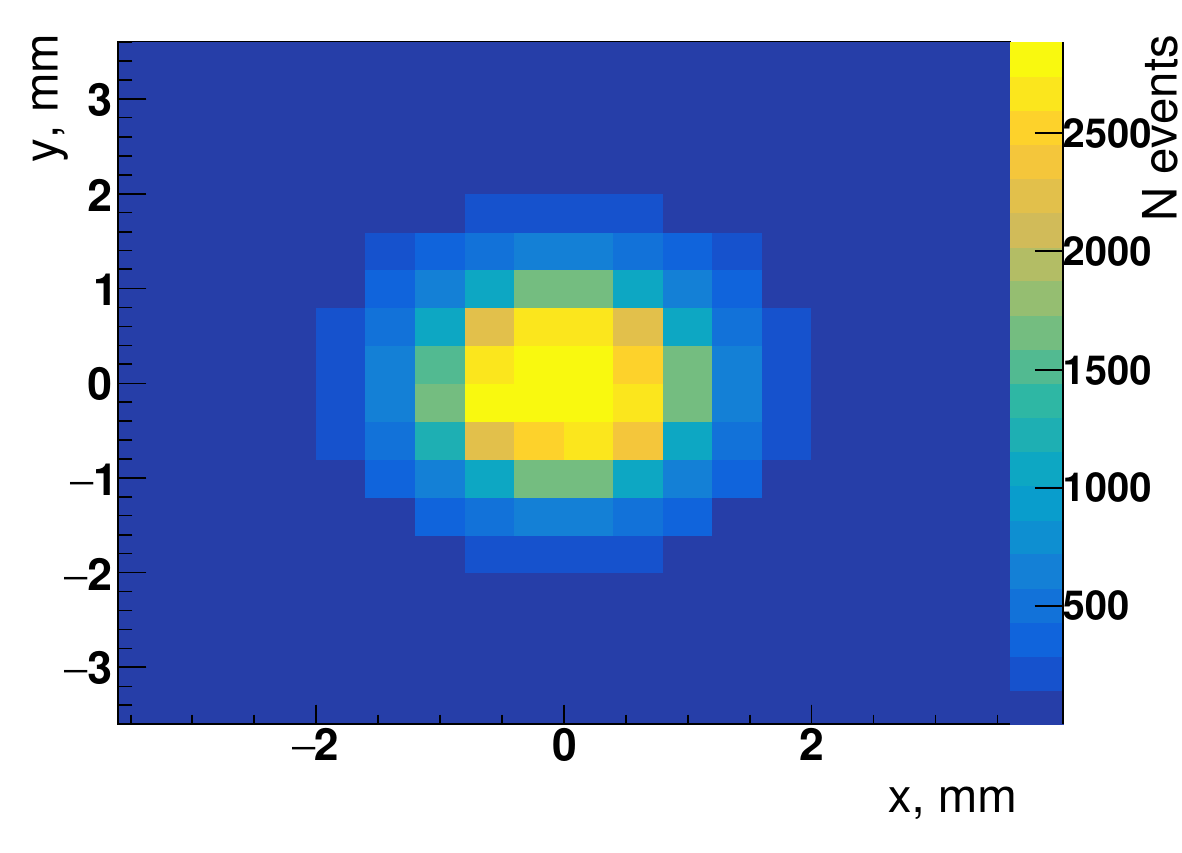}
    \caption{\label{fig:beam}
      Simulated spatial distribution of the $\gamma$ conversion vertex in the prototype.}
\vfill
    \centering\includegraphics[width=0.7\textwidth]{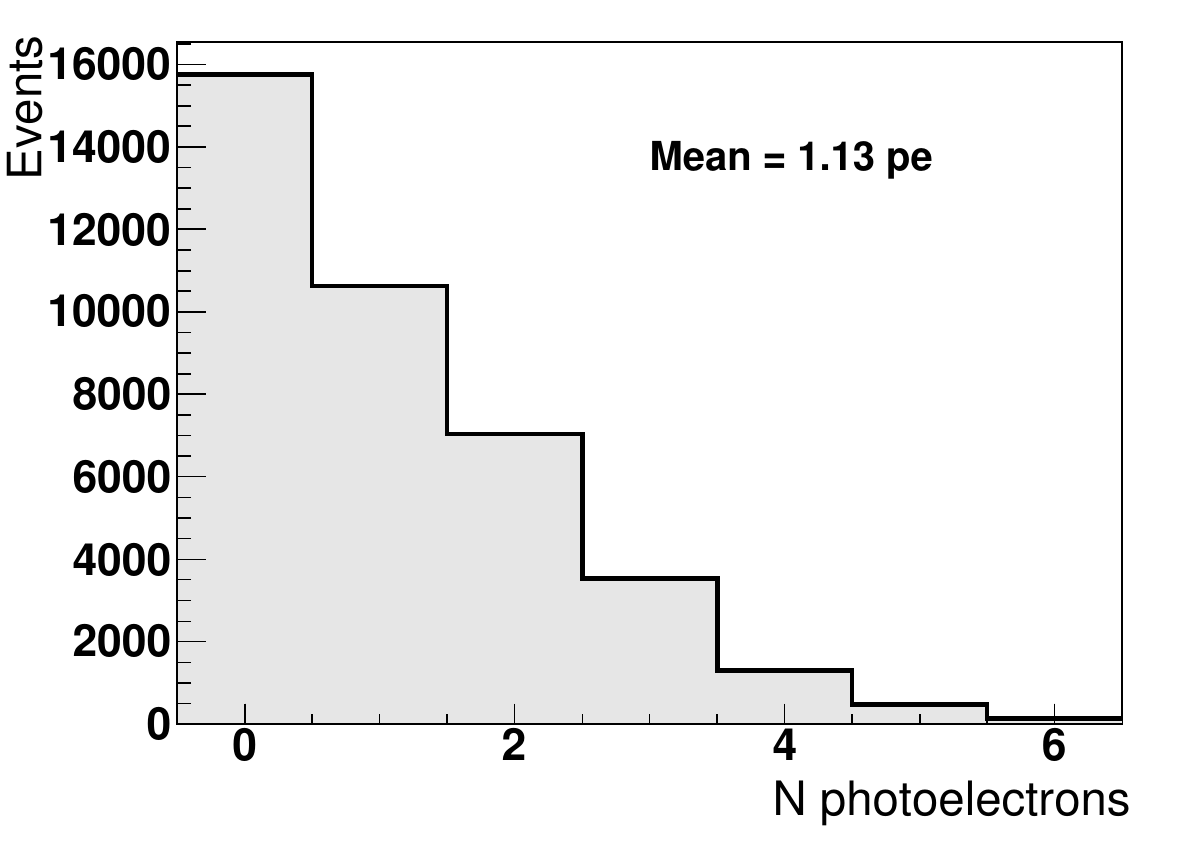}
    \caption{ \label{fig:npe}
      Number of detected photoelectrons per events with 511~keV
      energy deposition in the simulation. The mean number of photoelectrons is~1.1.
      }
  \end{minipage}
  \hfill
\begin{minipage}[ht]{0.38\columnwidth}
\centering\includegraphics[width=0.68\textwidth]{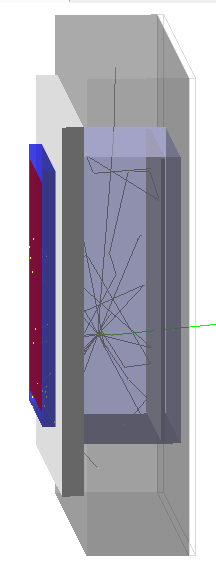}
\caption{\label{fig:simu}
  Visualization of the prototype implementation
  within the Geant4 simulation.}
\end{minipage}
\end{figure}

\begin{figure}[ht]
\centering
\includegraphics[width=.8\textwidth]{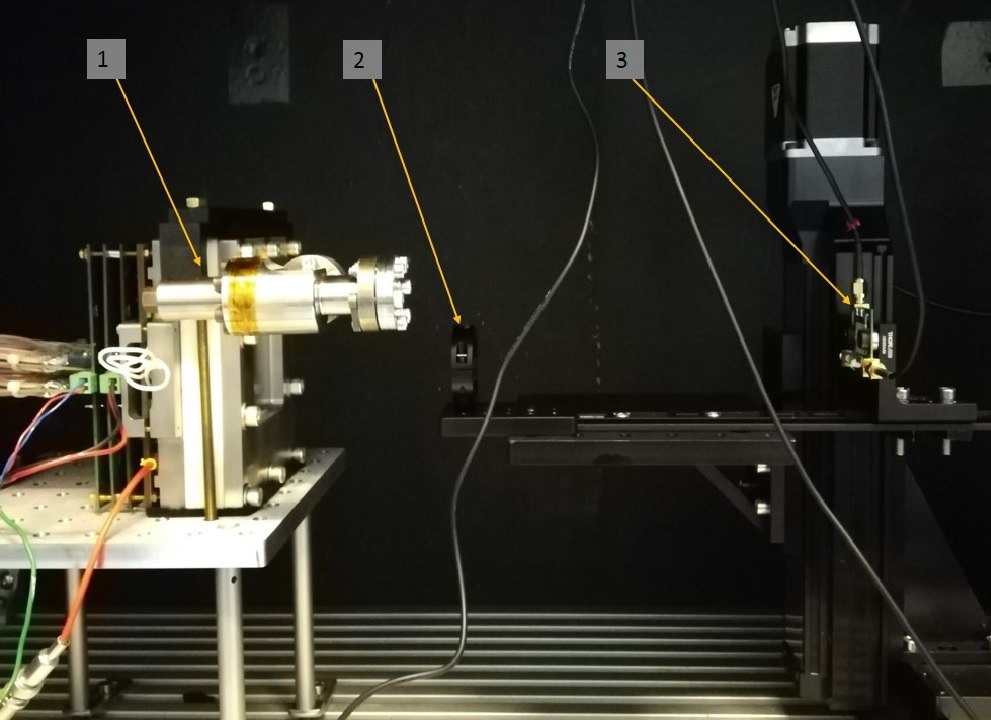}
\caption{\label{fig:coinc}
The coincidence setup consisting of the optical prototype (1), the $^{22}$Na  radioactive source (2), 
and the  LYSO 3x3x3 mm$^3$ spectrometer(3).}
\end{figure}

To evaluate the performance of the prototype in the PET-like  configuration, we employed a radioactive $^{22}$Na source.
This source emits positrons that promptly annihilates with electrons, 
resulting in pairs of 511~keV $\gamma$-photons emitted in opposite directions.
Simultaneously, a 1.275~MeV photon is released in a direction uncorrelated with those from the positron annihilation~\cite{PDG2022}.
The typical signals digitized by the SAMPIC modules are shown  in Fig.~\ref{fig:coincidence}.
To generate a well-defined electronically collimated beam of 511~keV photons, we positioned the source between the reference detector and the prototype. Data was acquired in coincidences between the two detectors.
When selecting events within the photoionization peak in the reference detector, we ensure that gamma rays are traveling directly from the source to the reference detector, as opposed to events outside the peak which
undergo the Compton scattering process and originate from different directions.
This allows for the formation of a well-defined beam in the prototype, with a size of approximately 2 mm (FWHM) according to
the simulations (section 2.4),~Fig.~\ref{fig:beam}.
This size limitation is due to the dimensions of the reference detector crystal, which measure 3 mm x 3 mm.
The radioactive source and reference detector are mounted on the 2D staging station and could be
moved to test different positions along the prototype surface. 
To minimize the impact of ambient light, all components were enclosed within a light-blocking black box, as shown in Fig.~\ref{fig:coinc}.

\subsection{Simulation}
\label{sec:simu}
In order to  understand the obtained results we developed a detailed simulation of the BOLDPET optical prototype~\cite{Ramos_2016,Kochebina2018Nov}. It is based on the Geant4 software \cite{agostinelli2003, Allison2006Feb, Allison2016Nov} and simulates
in full details the interaction of the $\gamma$-particles with matter, including the subsequent optical photons
generation and tracking.
In Fig.~\ref{fig:simu}, the visualization of the prototype configuration in the Geant4 simulation is shown.
The TMBi liquid appears in blue, the prototype's ceramic walls in grey, while the PMT window is depicted in dark blue
and the photocathode layer in red.
Individual optical photon trajectories within TMBi in a single event are visible as grey lines.
This simulation includes the following main parts of the detector response:
    \begin{itemize}
    \item $\gamma$ interaction in liquid TMBi accounts for three processes: photoelectric conversion,
      Compton scattering and Rayleigh diffusion. The first two processes produce relativistic electrons
      that emit visible photons through Cherenkov radiation process. 

    \item Each optical photon is propagated individually by the simulation program.
      During propagation, all key physical effects are considered: photon absorption within the TMBi liquid,
      reflection or absorption on the prototype walls, as well as on the optical windows and the optical gel.
      The refraction index of the TMBi changes between 1.58 (700 nm) and 1.63 (425 nm) as measured in~\cite{Ramos2015} with transparency 
      drops drastically  below 400~nm. We assume a diffusive reflection with an 86\% probability for ceramic reflectors,
      implemented using the Geant4 ``Unified Model''~\cite{Ramos_2016}.

        \item Bialkali photocathode simulation includes the Fresnel reflection of visible photons at photocathode boundaries,
          absorption of photons by the photocathode and extraction of generated photoelectrons as
          a function of the photon wavelength~\cite{Sung_2023}, Section~3.3.
        As a result, we compute that, on average, 1.1 photoelectrons is produced
        per 511~keV $\gamma$-ray photoelectric conversion in the liquid, Fig.~\ref{fig:npe}.

      \item We then parameterize the main PMT response features, including time response, PMT gain,
        gain fluctuation, and signal sharing between anodes, for more details, see~\cite{Sung_2023}, Section 3.4.

        \item Finally, we simulate the signal readout through the transmission lines with realistic signal shapes, taking into account the possible overlay of several photoelectrons, electronics noise and digitization sampling, ~\cite{Sung_2023}, Section 3.4.
              
    \end{itemize}

Most of the MCP-PMT simulation parameters (single-electron transition
time spread, PMT gain and gain fluctuation, signal sharing between
lines, etc. ) are adjusted to the results measured at the test bench
using pulsed laser in the single-photon regime~\cite{Follin2022Mar}.

Simulation results indicate a total detection efficiency $\varepsilon = \frac{N_{det}}{N_{total}} = 21.8\% \pm \%0.1 $ (stat),
where $N_{det}$ represents the number of detected events
and $N_{total}$ represents the total number of gamma rays passing through the prototype. 
The efficiency to detect events where photons deposit their energy entirely through photoionization conversion
($N_{ion}$) reaches $\varepsilon = \frac{N_{det}}{N_{ion}}= 56.9 \% \pm 0.2 \%$ (stat).

% section 3
\section {Results}

\subsection {Time resolution}
\label{sec:time}
\begin{figure}[ht]
%\begin{minipage}[ht]{0.49\textwidth}
  \centering\includegraphics[width=0.49\columnwidth]{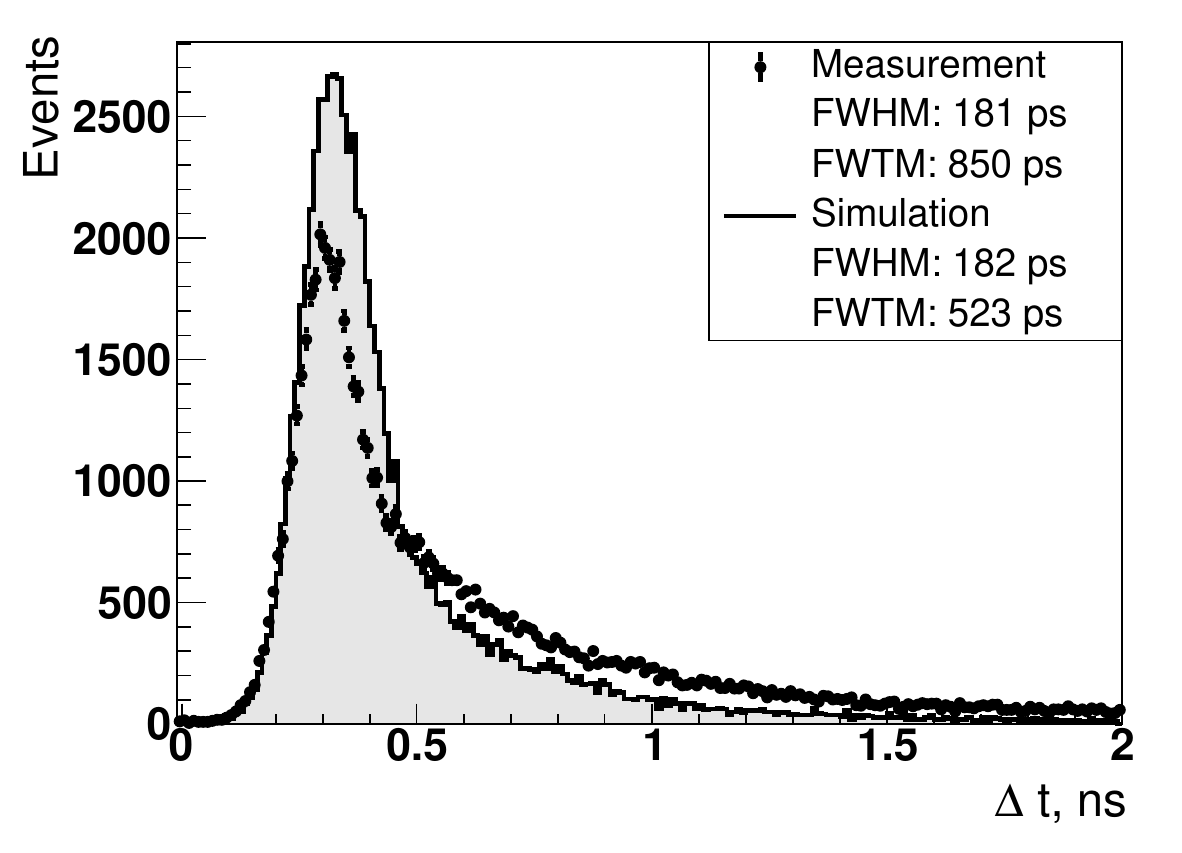}
  \caption{Measured (points with error bars)  and simulated (histogram) distributions of the time difference
    between the prototype and the reference detector. FWTM stands for Full Width at Tenth of Maximum.}
  \label{fig:time_reso}
%\end{minipage}
\end{figure}

To measure time resolution, we use a $^{22}$Na source, as detailed in Section~\ref{sec:source}. 
We acquire data in coincidence between the prototype and LYSO spectrometer. 
Given the intrinsic gain fluctuations of the MCP-PMT, we choose to estimate time using a constant fraction discrimination (CFD) algorithm
with a 50\% threshold~\cite{canot2018,Canot2019Dec}.
The signal time of the line is measured as $t = 0.5(t_R + t_L)$, where $t_R$ and $t_L$ represent the CFD times of the signal measured on the right and left sides of the line, respectively.
The PMT time is determined as the earliest time among all triggered lines\footnote{A "triggered line" refers to
  a  transmission line that has signals on both ends exceeding a specified threshold, typically, 50~mV.}.
The resolution is measured to be 180 ps (FWHM) with significant tail, as shown in Fig.\ref{fig:time_reso},
where the FWHM is calculated directly from the histograms without fitting the full histogram shape.
Our simulation predicts  the same resolution  with reduced tail. 
The difference between the experimental data and simulations is likely due to imperfections in modeling
the reflective properties of the wall surfaces.

\begin{figure}[ht]
\begin{subfigure}[t]{0.49\columnwidth}
    \centering\includegraphics[width=\textwidth]{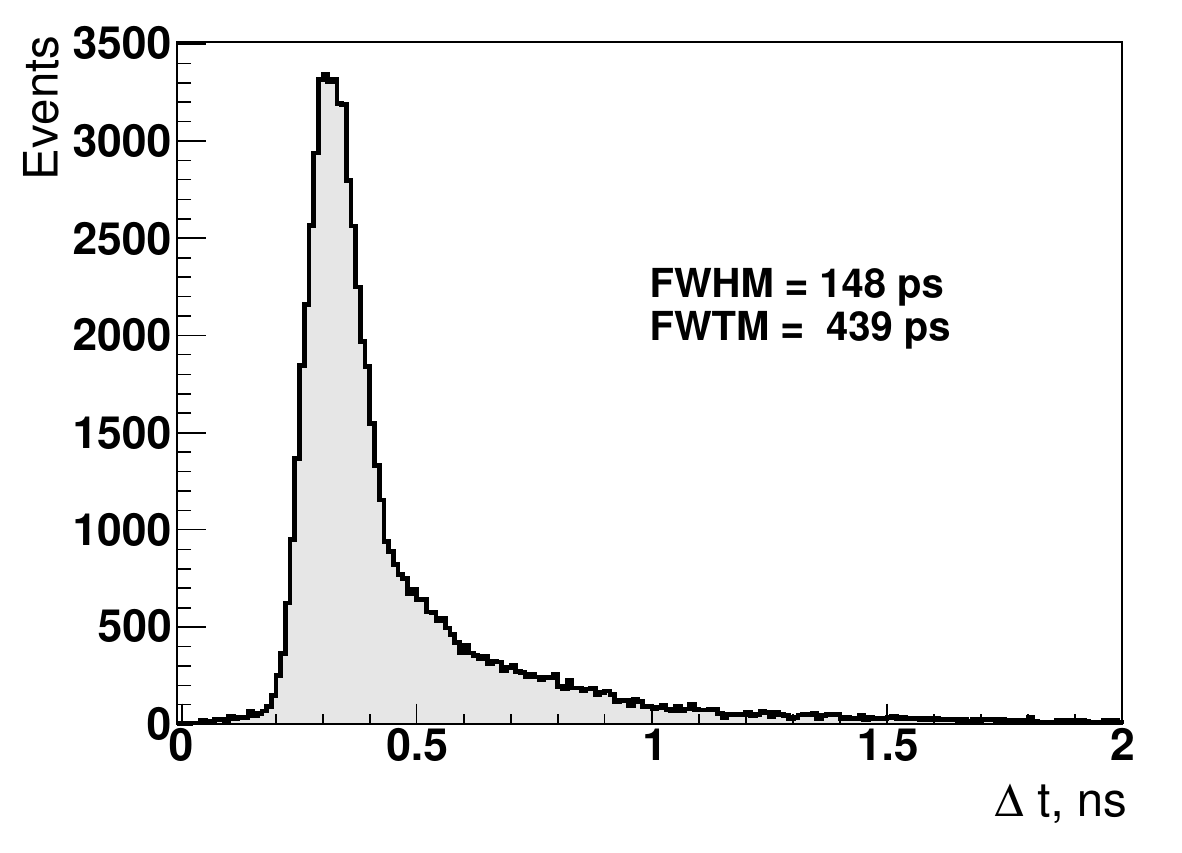}
    \caption{}
    \label{fig:time_reso_simu1a} 
\end{subfigure}
\hfill
\begin{subfigure}[t]{0.49\columnwidth}
    \centering\includegraphics[width=\textwidth]{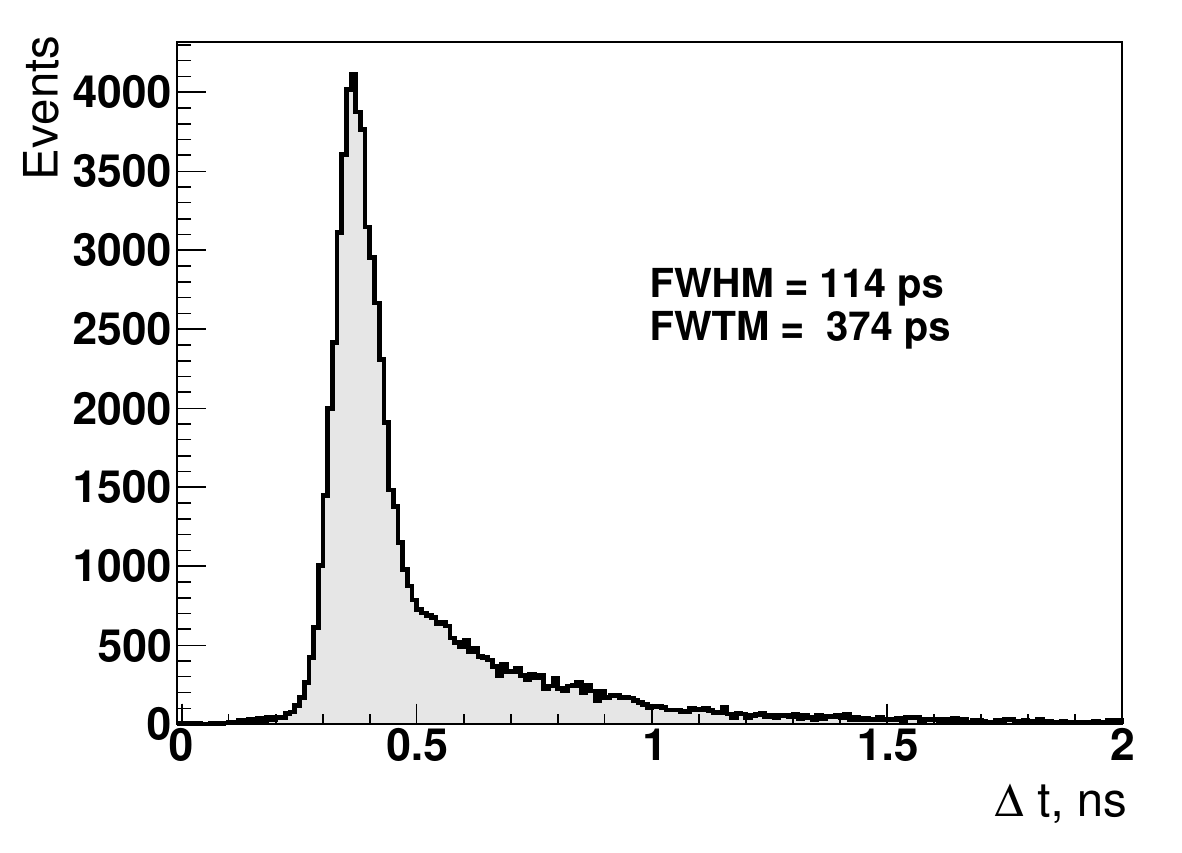}
    \caption{}
    \label{fig:time_reso_simu1b} 
\end{subfigure}
\caption{(a): the simulated distribution of the time difference between the prototype and the ideal reference detector with negligible time resolution. (b): the same, but corrected for the DOI coordinate (see text).}
\label{fig:time_reso_simu1} 
\end{figure}

\begin{figure}[ht]
\begin{subfigure}[t]{0.49\columnwidth}
    \centering\includegraphics[width=\columnwidth]{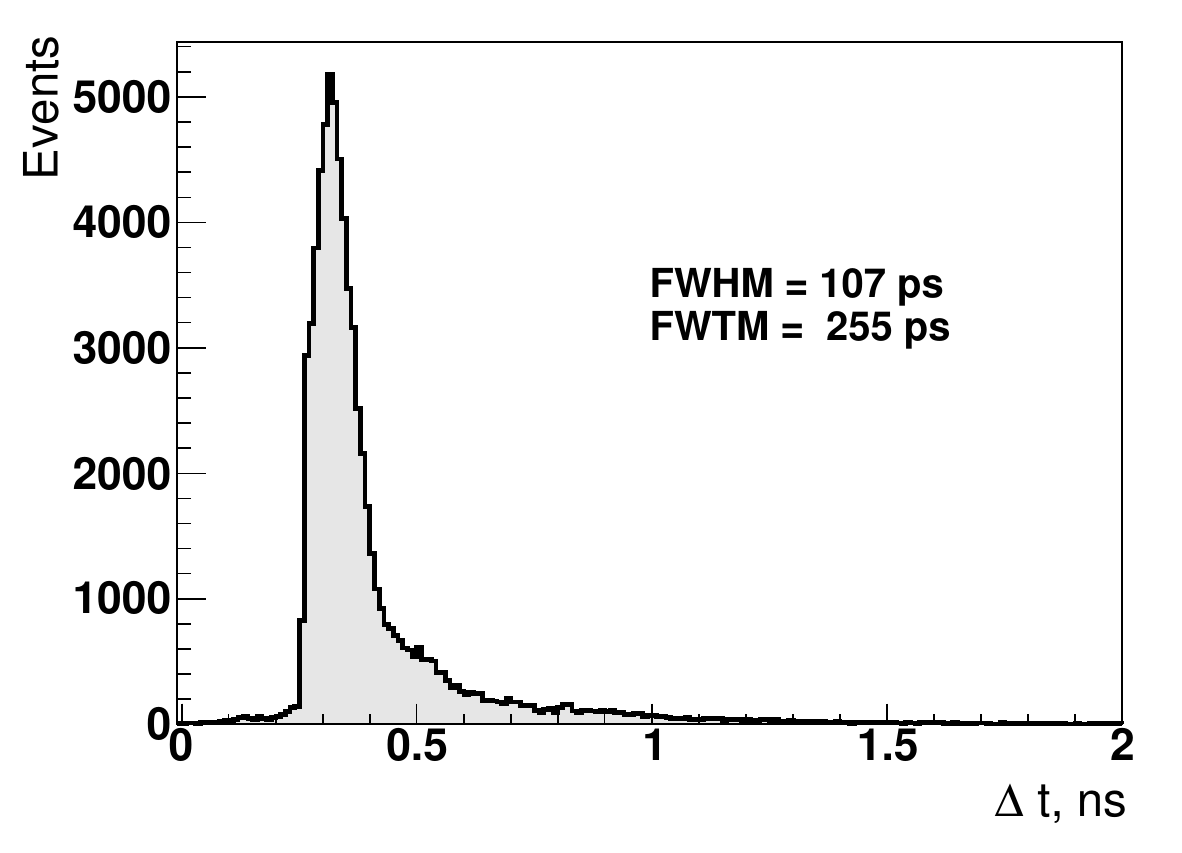}
    \caption{}
    \label{fig:time_reso_simu2a}
\end{subfigure}
\hfill
\begin{subfigure}[t]{0.49\columnwidth}
    \centering\includegraphics[width=\columnwidth]{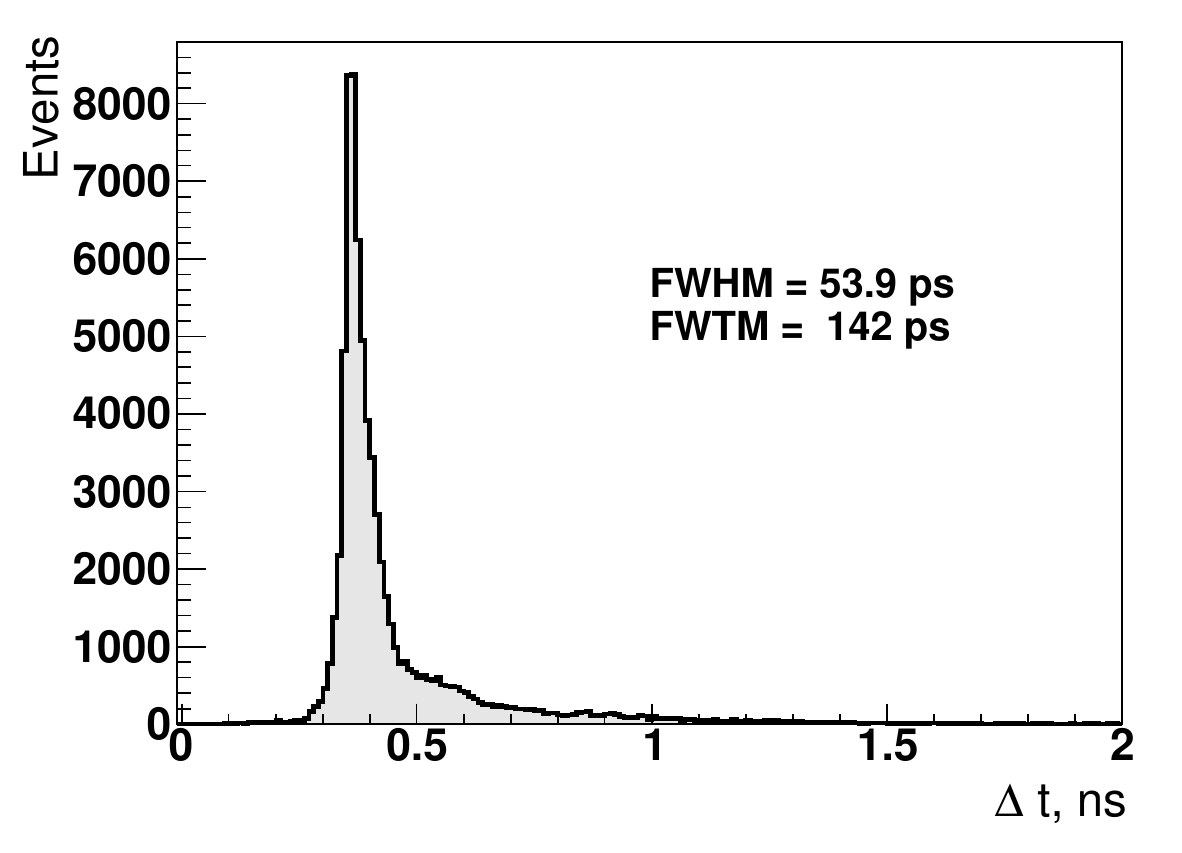}
    \caption{}
    \label{fig:time_reso_simu2b}
\end{subfigure}
\caption{(a): simulated distribution of the time difference between the prototype and  reference detector, assuming negligible time resolution for the reference detector and for the prototype PMT.
(b): The same, but corrected for the DOI coordinate (see text).}
\label{fig:time_reso_simu2} 
\end{figure}

Figure~\ref{fig:time_reso_simu1a} shows the simulated time difference between the prototype and ideal reference detector,
i.e., detector with negligible time resolution, thus demonstrating the  prototype time resolution of about 150~ps (FWHM). 
Further investigation into factors limiting time resolution is conducted through simulations assuming a negligible transit time spread\footnote{The PMT Transit Time Spread (TTS) refers to the variation in the time it takes for electrons to travel from the photocathode to the anodes and thereby characterizing the PMT time resolution.} (TTS) for the prototype PMT. The corresponding time difference distribution, Fig.~\ref{fig:time_reso_simu2a}, highlighted that the fluctuation of photon arrival times due to the dispersion of photon trajectories within the prototype  played a significant role in addition to the limitation imposed by the PMT TTS. To improve the resolution, one could consider reducing the thickness of the detection volume, at the cost of sacrificing $\gamma$ detection efficiency. 
Alternatively, the final BOLDPET detector design allows us to measure the depth of interaction (DOI) coordinate through the ionization signal. By exploiting the correlation between the mean photon arrival time and DOI, as elaborated in \cite{Ramos_2016}, we open another avenue for improvement. For instance, assuming that we are able to measure the
DOI coordinate with a precision of 5 mm (FWHM), we can significantly reduce the contribution due to fluctuations in photon collection time, thereby improving the time resolution from 107 ps to a remarkable 54 ps, as illustrated in Fig.\ref{fig:time_reso_simu2b}. Implementing the same correction in the prototype with realistic PMT TTS will improve its resolution from 148 ps to a noteworthy 114 ps, as demonstrated in Fig.\ref{fig:time_reso_simu1b}.

%% \begin{figure}[ht]
%% \begin{subfigure}[t]{0.49\columnwidth}
%%     \centering\includegraphics[width=\columnwidth]{h_dt.eps}
%%     \caption{\label{fig:time_reso_meas}}
%% \end{subfigure}
%% \hfill
%% \begin{subfigure}[t]{0.49\columnwidth}
%% \centering\includegraphics[width=\columnwidth]{h_dt_simu.eps}
%% \caption{\label{fig:time_reso_simu}} 
%% \end{subfigure}
%% \caption{Measured (a)  and simulated (b) distributions of the time difference between the prototype and the reference detector.
%% FWTM stands for Full Width at Tenth of Maximum.}
%% \label{fig:time_reso}
%% \end{figure}

%subsection 4.2
\subsection {Spatial resolution}
\label{sec:spatial}
The optical part of the BOLDPET detection module does not aim to provide high spatial resolution, since the ionization signal would be used to accurately measure three spatial coordinates. At the same time, the ability to measure the interaction position will help to resolve the ambiguity when multiple interaction vertices exist within the volume. This, in turn, leads to an enhancement in the maximum count rate supported by the detector.

In Section~\ref{sec:time}, we discuss the data collected for time measurements, which is also employed to evaluate spatial resolution. The requirement for coincidence between the prototype and the reference detector leads to a 2~mm wide $\gamma$-beam in the former, as described in the section~\ref{sec:source} and Fig.~\ref{fig:beam}. 

To determine the coordinate along the transmission lines,
the x-coordinate, we reconstruct it using the line with the highest amplitude, $i$.  
The $\gamma$-interaction x-coordinate is determined as the time difference
between signals read out from the right and left sides of the line,
$t_R$ and $t_L$ respectively, multiplied by the signal propagation speed, $s$,
and divided by two, i.e. $x_\gamma = ( t_R^i-t_L^i ) s/2$.
The signal propagation speed is measured to be about 0.36 times speed-of-light.

For the coordinate across the lines, the y-coordinate, we use the line with the highest amplitude, denoted as $i$, and its two neighboring lines. We compute the charge-weighted coordinate as follows:
\begin{equation}
    y_\gamma=\frac{\sum _{k=i-1}^{i+1}y_k C_k}{\sum _{k=i-1}^{i+1} C_k}
\end{equation}
where $y_k$ corresponds to the coordinate of the center of line $k$, and $C_k$ corresponds to its charge.

\begin{figure}[ht]
\begin{subfigure}[t]{\columnwidth}
    \centering\includegraphics[width=0.49\columnwidth]{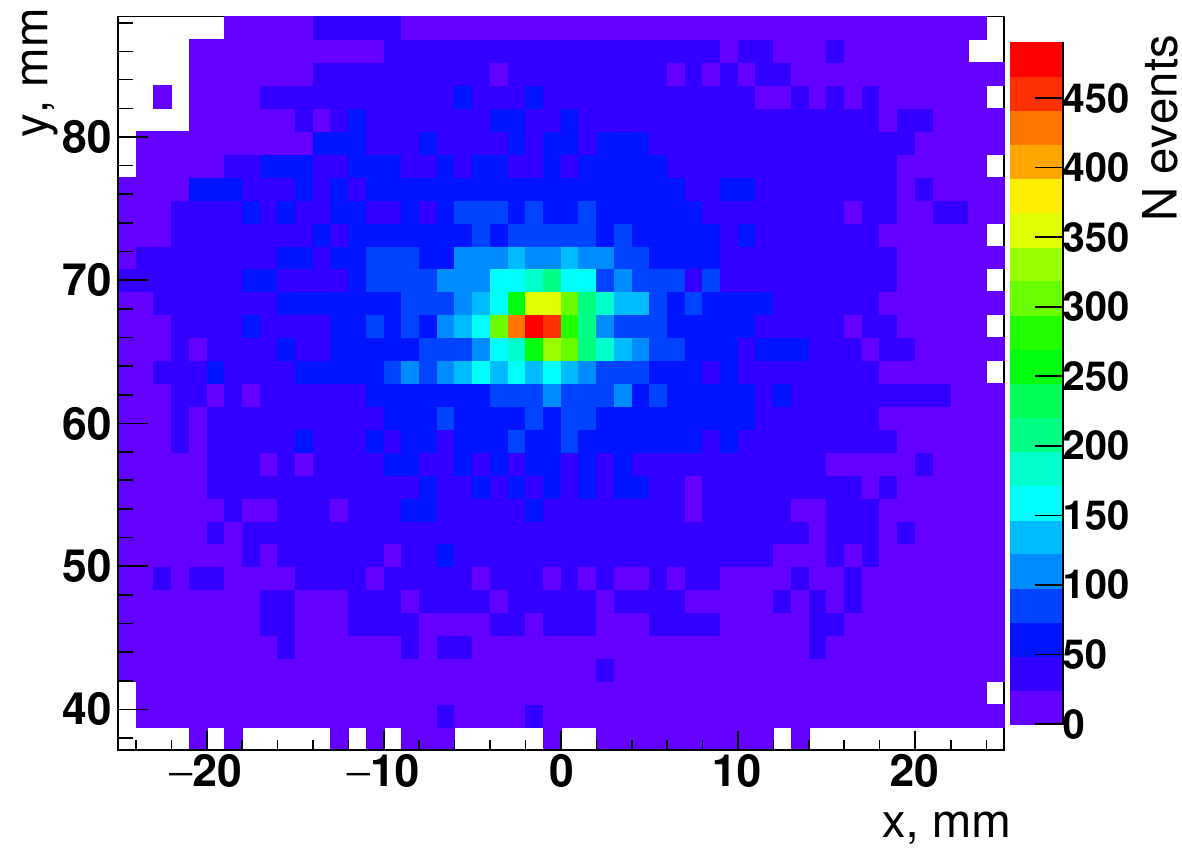}
\hfill
    \centering\includegraphics[width=0.49\columnwidth]{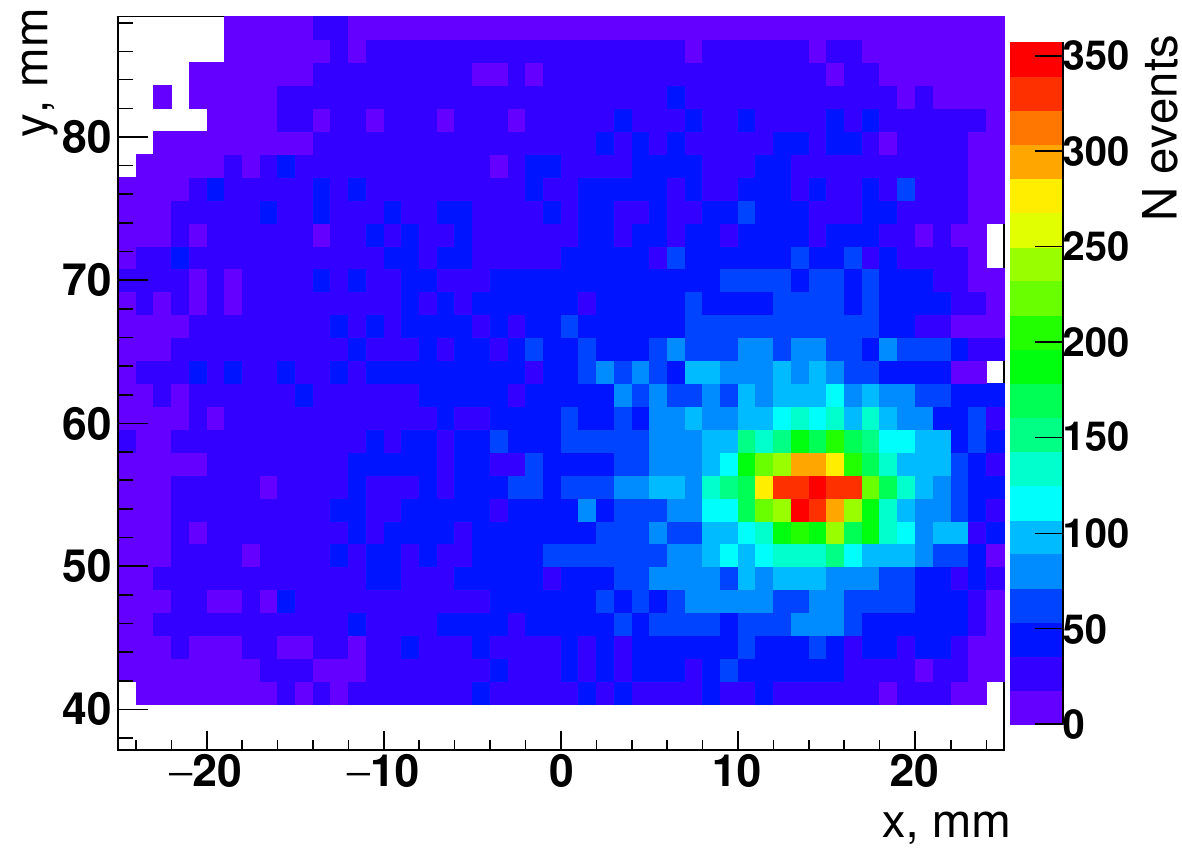}
\caption{(x,y) 2d-distributions}
    \label{fig:2dreco}
\end{subfigure}

\begin{subfigure}[t]{\columnwidth}
    \centering\includegraphics[width=0.49\columnwidth]{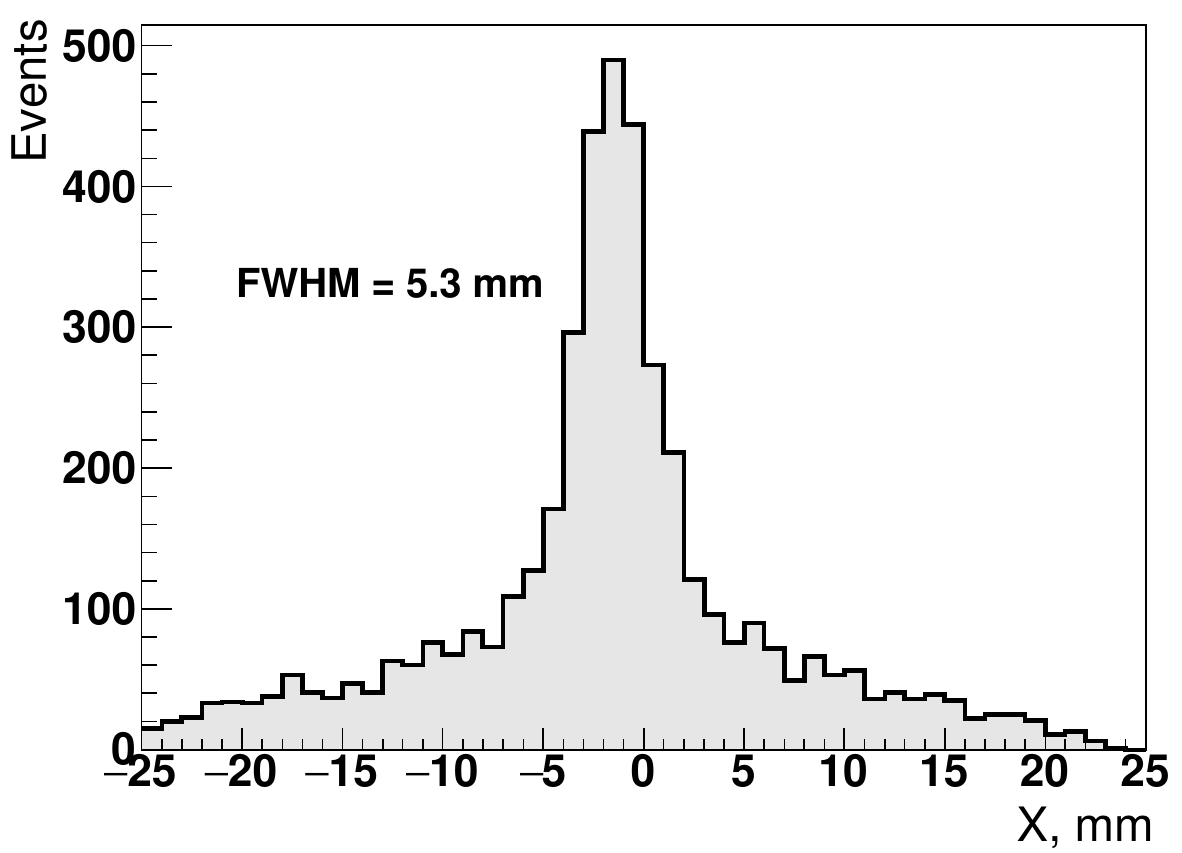}
\hfill
    \centering\includegraphics[width=0.49\columnwidth]{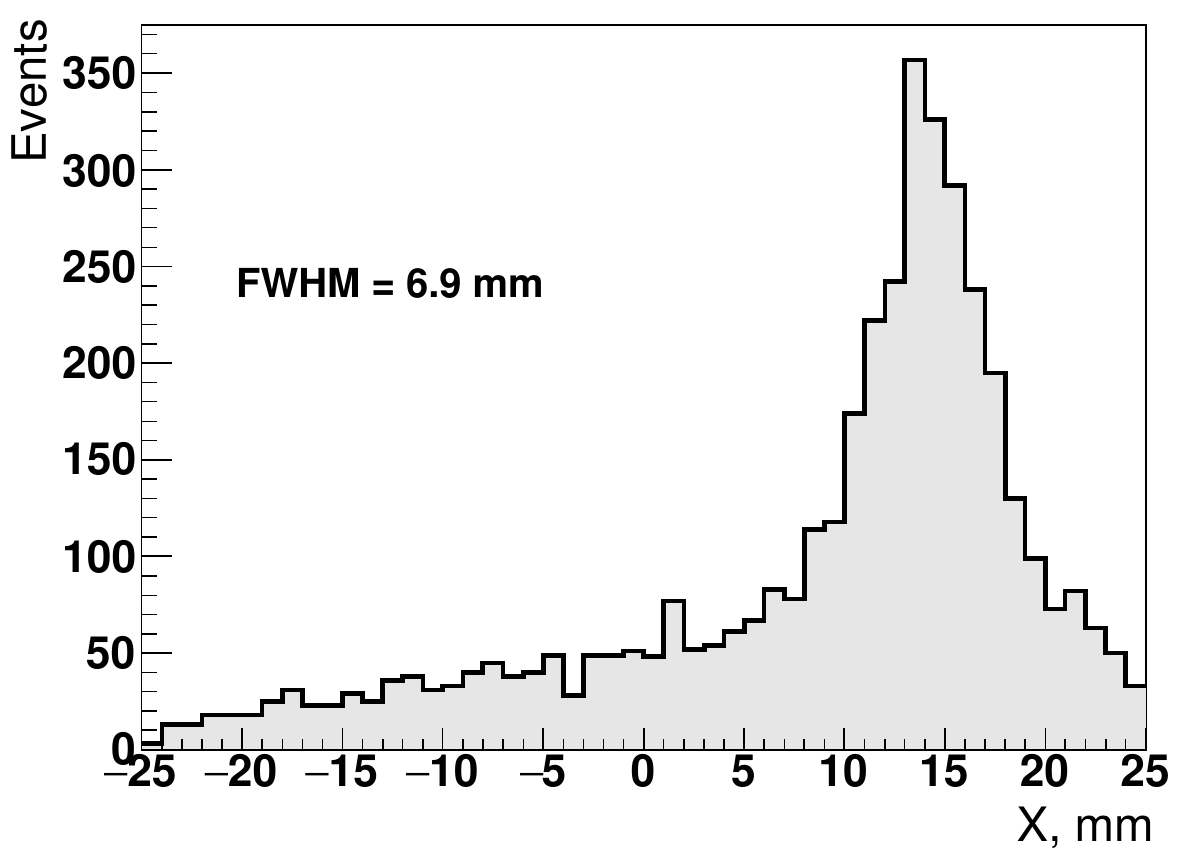}
\caption{x-coordinate distribution}
\label{fig:xmax}
\end{subfigure}

\begin{subfigure}[t]{\columnwidth}
    \centering\includegraphics[width=0.49\columnwidth]{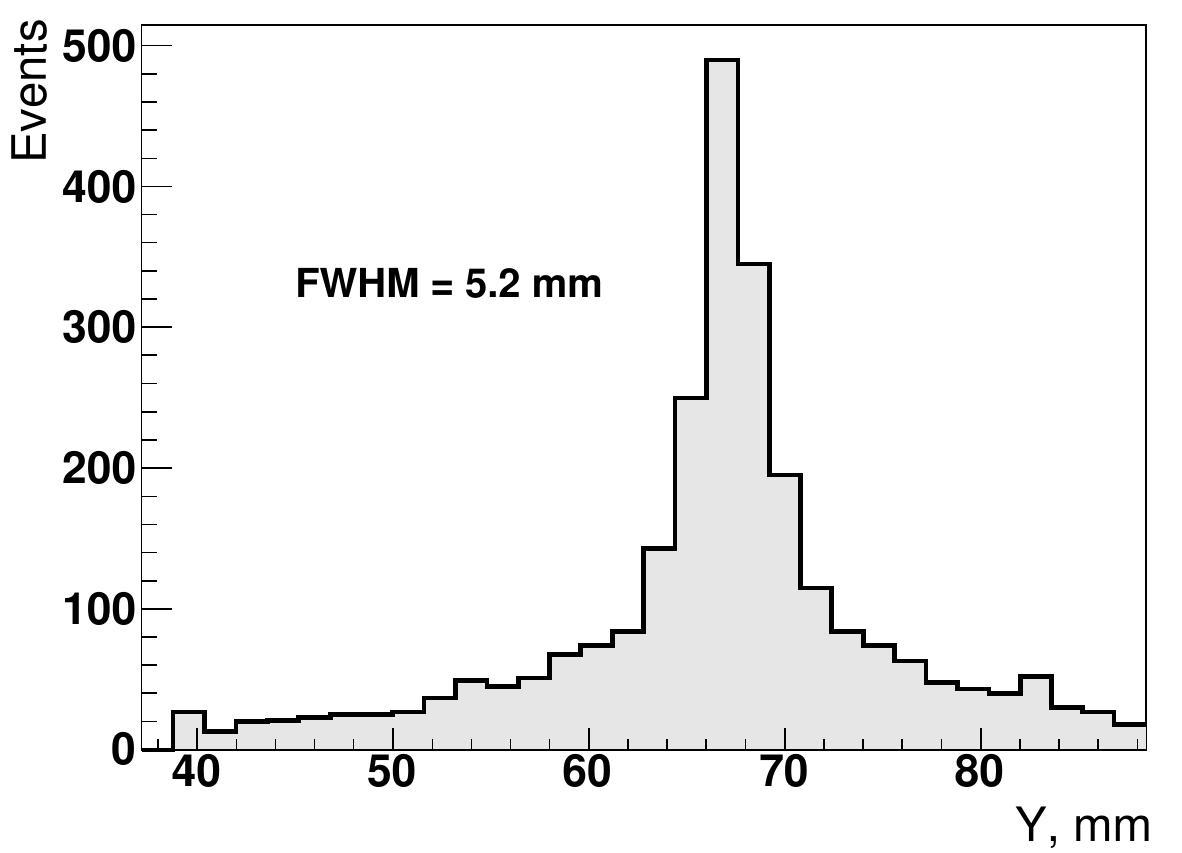}
\hfill
    \centering\includegraphics[width=0.49\columnwidth]{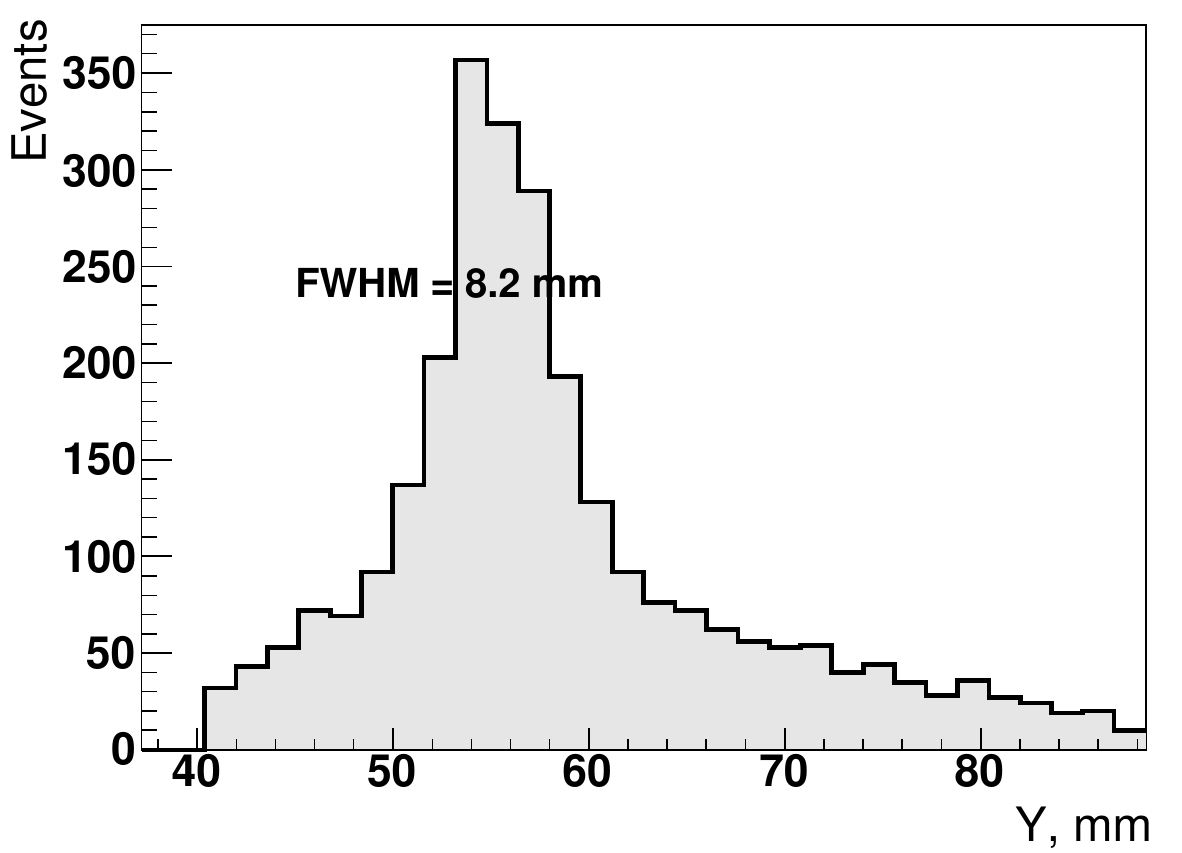}
    \caption{y-coordinate distribution}
    \label{fig:ymax2}
\end{subfigure}

\caption{Reconstructed positions of the $\gamma$-interaction vertices in the prototype for 
two distinct positions of the $^{22}$Na $\gamma$-beam,
shown in the left and right columns, respectively.
The 1D distributions represent the x and y cross-sections of the 2D histograms taken at the highest bin of the distribution
on the orthogonal axis.
}
\label{fig:reco_coord} 
\end{figure}

Figure~\ref{fig:2dreco} displays the 2D $(x, y)$ distributions of the reconstructed $\gamma$-interaction vertices for two different positions of the $\gamma$ beam. The $x$ and $y$ cross-sections reveal a resolution width of approximately 5~mm for positions near the center of the detector (left column), and 7~mm to 8~mm for positions in the corner of the detector.
This level of resolution will be valuable in resolving ambiguity when associating the optical signal with ionization in the case of multiple vertices in the detection volume.

% section 5
\section{Discussions}

The current work highlights the significant potential of using
Cherenkov photons to enhance the time resolution when detecting
511~keV photons. The developed detector technique has the advantage of
producing optical signals due to the Cherenkov photons only.
As mentioned in Section~\ref{sec:simu} the efficiency to detect events where photons deposit their energy entirely
through photoionization conversion is about 57\%.
This value is constrained by the
limited number of produced Cherenkov photons, of about 20 for a
511~keV energy deposition. Consequently the mean number of
photo-electrons detected by the PMT is approximately 1.1, as
illustrated in Fig.~\ref{fig:npe}.

\begin{figure}[ht]
\begin{subfigure}[t]{0.49\columnwidth}
    \centering\includegraphics[width=\columnwidth]{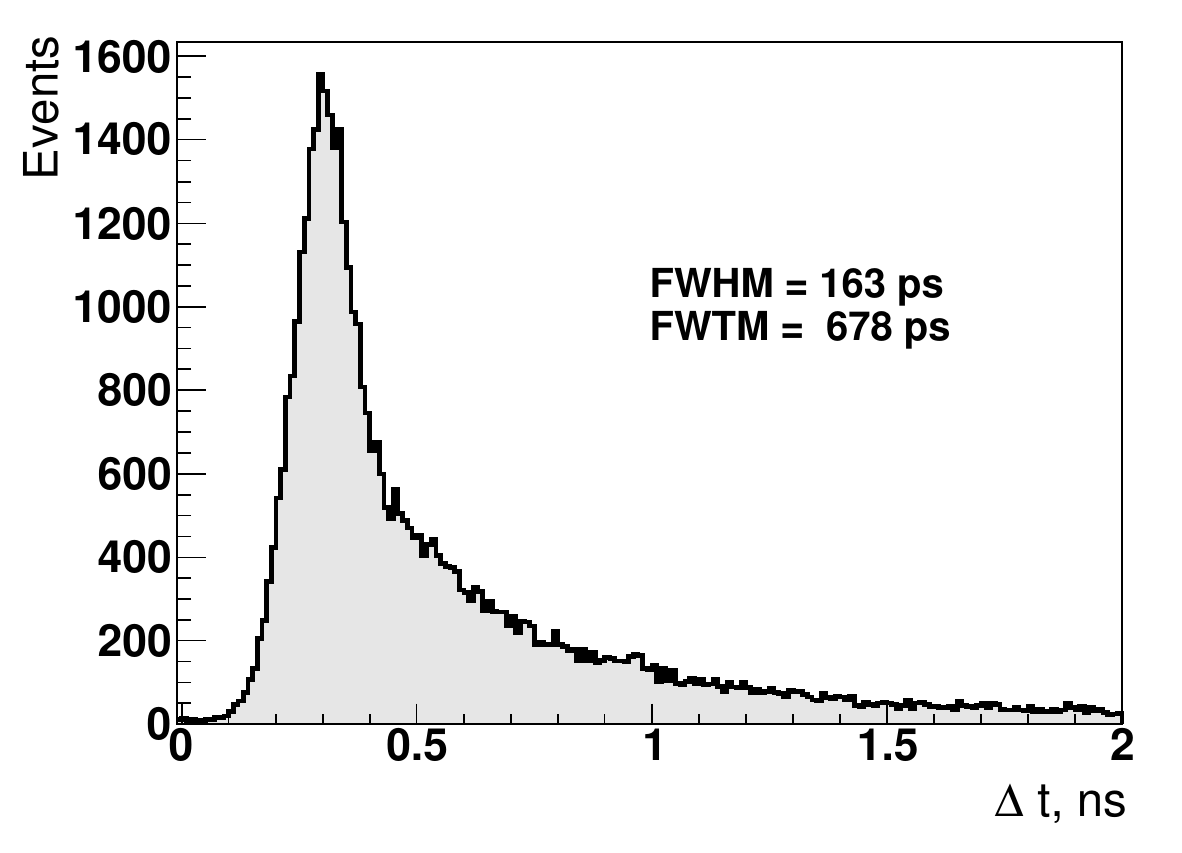}
   \caption{2 or more triggered lines } 
\end{subfigure}
\hfill
\begin{subfigure}[t]{0.49\columnwidth}
    \centering\includegraphics[width=\columnwidth]{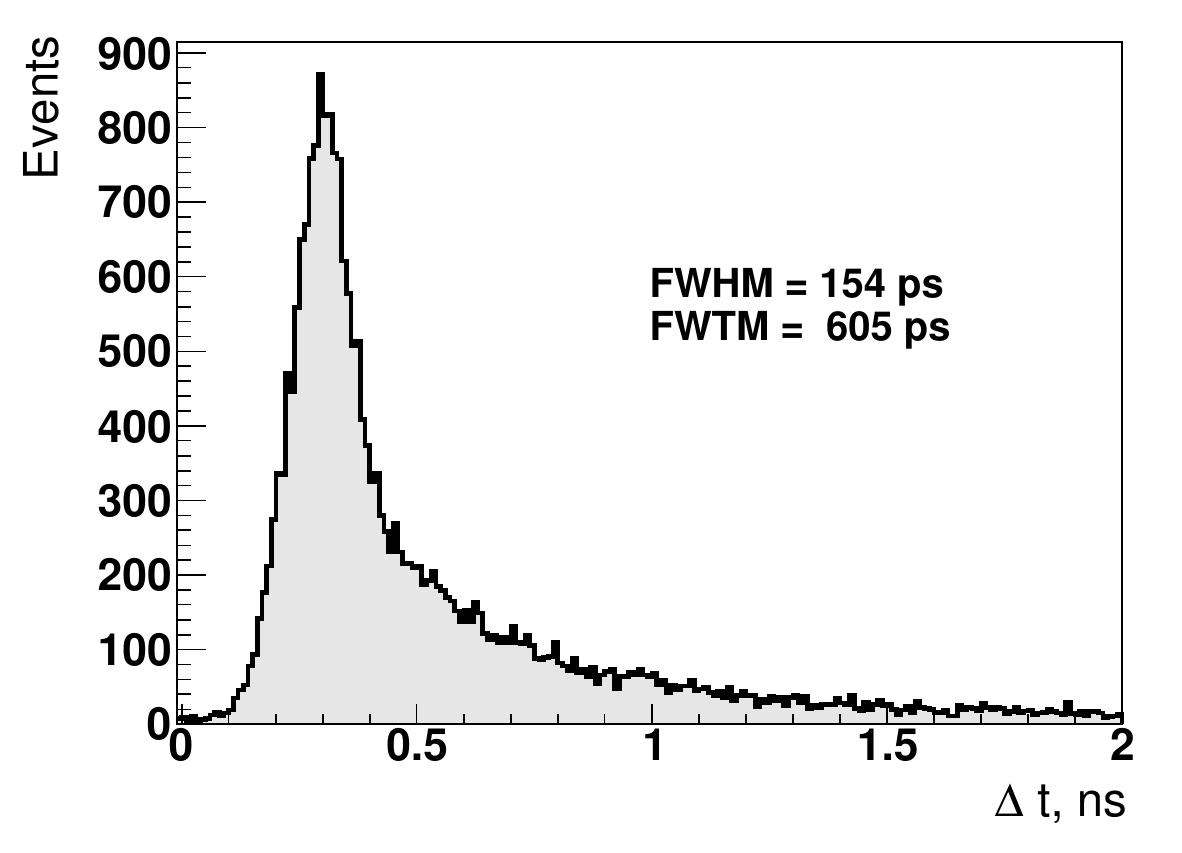}
\caption{3 or more triggered lines }
\end{subfigure}
\caption{Time difference distribution between the prototype and the LYSO spectrometer 
considering events with more than one triggered lines.}
\label{fig:dt_nlines} 
\end{figure}

As discussed in Section~\ref{sec:time}, the time resolution faces two
key limitations: fluctuations in the optical photon path and the
limited precision of available photodetectors.  Mitigating the first
factor involves optimizing the detector thickness to balance detection
efficiency and time resolution.  An additional strategy involves
measuring DOI coordinate and correcting for the
correlation between mean detection time and DOI.  Addressing the
second factor requires increasing photon detection efficiency for an
overall enhancement in time resolution. For instance,
Fig.~\ref{fig:dt_nlines} shows the measured time resolution in events
with more than one triggered line

These distributions reveal a substantial decrease in precision, down
to 154~ps, corresponding to a detector resolution of about 120~ps
(FWHM). This occurs because a larger number of triggered lines
correlates with a higher quantity of produced photoelectrons.

Looking ahead, we may hope for the improvement in the PMT time
response function thereby enhancing overall time performance. This is
particularly promising, given that photons produced through the
Cherenkov mechanism are generated on the timescale of several
picoseconds. Therefore, any improvement in PMT time resolution
directly translates to an enhanced $\gamma$-detection time resolution
achieved through the detection of Cherenkov photons.

Although these conclusions are drawn from experiences with the BOLDPET
optical prototype, they hold general applicability to any utilization
of Cherenkov radiation in PET detection, such as employing it in
crystalline radiators.

% section 6 (Conclusions)
\section{Conclusions}
In this work we investigate the performance of the optical prototype
developed for the BOLDPET project in detecting 511~keV
$\gamma$-quanta. The unique feature of this detection technology is
the utilization of the heavy detection liquid, TMBi, enabling the
design of large-scale detection modules with a high detection
efficiency.  The generation of optical photons occurs exclusively
through the Cherenkov mechanism, and they are detected by fast
micro-channel plate photo-multiplier, resulting in an excellent time
resolution for the large scale detectors of approximately 150~ps
(FWHM).  Further refinement of this resolution to 120~ps is achievable
by implementing additional selection criteria, albeit at the expense
of reduced efficiency.  We demonstrate the feasibility of coarse
localization of interactions using solely the optical signal,
achieving a precision of about 5-8 mm (FWHM).  Using detailed Geant4
simulations, we explore the limiting factors influencing time
resolution and discuss potential avenues for improvement.

\section*{Acknowledgments}
The authors would like to thank Saint-Gobain Crystals for kindly providing the LYSO:Ce:Ca single crystals used to build LYSO spectrometers.\\
We acknowledge the financial support by the joint French-German grants  ANR-18-CE92-0012-01, DFG-SCHA 1447/3-1 and WE 1843/8-1.

\bibliography{mybib}

\providecommand{\href}[2]{#2}\begingroup\raggedright\begin{thebibliography}{10}

\bibitem{Vallabhajosula_2011}
S.~Vallabhajosula, L.~Solnes and B.~Vallabhajosula, \emph{{A Broad Overview of
  Positron Emission Tomography Radiopharmaceuticals and Clinical Applications:
  What Is New?}},
  \href{https://doi.org/10.1053/j.semnuclmed.2011.02.003}{\emph{Semin. Nucl.
  Med.} {\bfseries 41} (2011) 246}.

\bibitem{Martinez2019Jun}
O.~Martinez, J.~Sosabowski, J.~Maher and S.~Papa, \emph{{New Developments in
  Imaging Cell-Based Therapy}},
  \href{https://doi.org/10.2967/jnumed.118.213348}{\emph{Journal of Nuclear
  Medicine} {\bfseries 60} (2019) 730}.

\bibitem{Djekidel2022Sep}
M.~Djekidel, R.~AlSadi, M.A.~Akl, S.~Vandenberghe and O.~Bouhali,
  \emph{{Total-body pediatric PET is ready for prime time}},
  \href{https://doi.org/10.1007/s00259-022-05873-y}{\emph{European Journal of
  Nuclear Medicine and Molecular Imaging} {\bfseries 49} (2022) 3624}.

\bibitem{HRRT_deJong_2007}
H.W.A.M.~de~Jong, F.H.P.~van Velden, R.W.~Kloet, F.L.~Buijs, R.~Boellaard and
  A.A.~Lammertsma, \emph{{Performance evaluation of the ECAT HRRT: an LSO-LYSO
  double layer high resolution, high sensitivity scanner}},
  \href{https://doi.org/10.1088/0031-9155/52/5/019}{\emph{Phys. Med. Biol.}
  {\bfseries 52} (2007) 1505}.

\bibitem{Gonzalez2018Jun}
A.J.~Gonz{\ifmmode\acute{a}\else\'{a}\fi}lez,
  F.~S{\ifmmode\acute{a}\else\'{a}\fi}nchez and J.M.~Benlloch,
  \emph{{Organ-Dedicated Molecular Imaging Systems}},
  \href{https://doi.org/10.1109/TRPMS.2018.2846745}{\emph{IEEE Transactions on
  Radiation and Plasma Medical Sciences} {\bfseries 2} (2018) 388}.

\bibitem{Vandenberghe2020Dec}
S.~Vandenberghe, P.~Moskal and J.S.~Karp, \emph{{State of the art in total body
  PET}}, \href{https://doi.org/10.1186/s40658-020-00290-2}{\emph{EJNMMI
  Physics} {\bfseries 7} (2020) 1}.

\bibitem{Zeimpekis2022Jul}
K.G.~Zeimpekis, F.A.~Kotasidis, M.~Huellner, A.~Nemirovsky, P.A.~Kaufmann and
  V.~Treyer, \emph{{NEMA NU 2{\textendash}2018 performance evaluation of a new
  generation 30-cm axial field-of-view Discovery MI PET/CT}},
  \href{https://doi.org/10.1007/s00259-022-05751-7}{\emph{European Journal of
  Nuclear Medicine and Molecular Imaging} {\bfseries 49} (2022) 3023}.

\bibitem{Saanchez-Crespo2004}
A.~Saanchez-Crespo, P.~Andreo and S.A.~Larsson, \emph{{Positron flight in human
  tissues and its influence on {PET} image spatial resolution}},
  \href{https://doi.org/10.1007/s00259-003-1330-y}{\emph{Eur. J. Nucl. Med.
  Mol. Imaging} {\bfseries 31} (2004) 44}.

\bibitem{Harpen2004Jan}
M.D.~Harpen, \emph{{Positronium: Review of symmetry, conserved quantities and
  decay for the radiological physicist}},
  \href{https://doi.org/10.1118/1.1630494}{\emph{Med. Phys.} {\bfseries 31}
  (2004) 57}.

\bibitem{Moses2011a}
W.W.~Moses, \emph{{Fundamental limits of spatial resolution in PET}},
  \href{https://doi.org/10.1016/j.nima.2010.11.092}{\emph{Nucl. Instrum. Meth.
  A} {\bfseries 648} (2011) S236 }.

\bibitem{Lehnert2011May}
W.~Lehnert, M.-C.~Gregoire, A.~Reilhac and S.R.~Meikle, \emph{{Analytical
  positron range modelling in heterogeneous media for PET Monte Carlo
  simulation}},
  \href{https://doi.org/10.1088/0031-9155/56/11/009}{\emph{Physics in Medicine
  {\&} Biology} {\bfseries 56} (2011) 3313}.

\bibitem{Emond2019Oct}
E.C.~Emond, A.M.~Groves, B.F.~Hutton and K.~Thielemans, \emph{{Effect of
  positron range on PET quantification in diseased and normal lungs}},
  \href{https://doi.org/10.1088/1361-6560/ab469d}{\emph{Phys. Med. Biol.}
  {\bfseries 64} (2019) 205010}.

\bibitem{Belcari2017Jun}
N.~Belcari, N.~Camarlinghi, S.~Ferretti, P.~Iozzo, D.~Panetta, P.A.~Salvadori
  et~al., \emph{{NEMA NU-4 Performance Evaluation of the IRIS PET/CT
  Preclinical Scanner}},
  \href{https://doi.org/110.1109/TRPMS.2017.2707300}{\emph{IEEE Transactions on
  Radiation and Plasma Medical Sciences} {\bfseries 1} (2017) 301}.

\bibitem{Amirrashedi2019Nov}
M.~Amirrashedi, S.~Sarkar, P.~Ghafarian, R.H.~Shahraki, P.~Geramifar, H.~Zaidi
  et~al., \emph{{NEMA NU-4 2008 performance evaluation of Xtrim-PET: A
  prototype SiPM-based preclinical scanner}},
  \href{https://doi.org/10.1002/mp.13785}{\emph{Medical Physics} {\bfseries 46}
  (2019) 4816}.

\bibitem{Gaudin2021Mar}
{\ifmmode\acute{E}\else\'{E}\fi}.~Gaudin, C.~Thibaudeau, L.~Arpin,
  J.-D.~Leroux, M.~Toussaint, J.-F.~Beaudoin et~al., \emph{{Performance
  evaluation of the mouse version of the LabPET II PET scanner}},
  \href{https://doi.org/10.1088/1361-6560/abd952}{\emph{Physics in Medicine
  {\&} Biology} {\bfseries 66} (2021) 065019}.

\bibitem{Vandenberghe_2016}
S.~Vandenberghe, E.~Mikhaylova, E.~D'Hoe, P.~Mollet and J.S.~Karp,
  \emph{{Recent developments in time-of-flight {PET}}},
  \href{https://doi.org/10.1186/s40658-016-0138-3}{\emph{{EJNMMI} Physics}
  {\bfseries 3} (2016) }.

\bibitem{Lecoq2020Oct}
P.~Lecoq, C.~Morel, J.O.~Prior, D.~Visvikis, S.~Gundacker, E.~Auffray et~al.,
  \emph{{Roadmap toward the 10 ps time-of-flight PET challenge}},
  \href{https://doi.org/10.1088/1361-6560/ab9500}{\emph{Physics in Medicine
  {\&} Biology} {\bfseries 65} (2020) 21RM01}.

\bibitem{Schaart2021Apr}
D.R.~Schaart, \emph{{Physics and technology of time-of-flight PET detectors}},
  \href{https://doi.org/10.1088/1361-6560/abee56}{\emph{Physics in Medicine
  {\&} Biology} {\bfseries 66} (2021) 09TR01}.

\bibitem{Mohammadi2019}
I.~Mohammadi, I.F.C.~Castro, P.M.M.~Correia, A.L.M.~Silva and J.F.C.A.~Veloso,
  \emph{{Minimization of parallax error in positron emission tomography using
  depth of interaction capable detectors: methods and apparatus}},
  \href{https://doi.org/10.1088/2057-1976/ab4a1b}{\emph{Biomed. Phys. Eng.
  Express} {\bfseries 5} 062001}.

\bibitem{Borghi_2016b}
G.~Borghi, B.J.~Peet, V.~Tabacchini and D.R.~Schaart, \emph{{A 32 mm x 32 mm x
  22 mm monolithic {LYSO}:Ce detector with dual-sided digital photon counter
  readout for ultrahigh-performance {TOF}-{PET} and {TOF}-{PET}/{MRI}}},
  \href{https://doi.org/10.1088/0031-9155/61/13/4929}{\emph{Phys. Med. Biol.}
  {\bfseries 61} (2016) 4929}.

\bibitem{Gonzalez-Montoro2017}
A.~Gonzalez-Montoro, A.~Aguilar, G.~Canizares, P.~Conde, L.~Hernandez,
  L.F.~Vidal et~al., \emph{{Performance Study of a Large Monolithic LYSO PET
  Detector With Accurate Photon DOI Using Retroreflector Layers}},
  \href{https://doi.org/10.1109/TRPMS.2017.2692819}{\emph{IEEE Transactions on
  Radiation and Plasma Medical Sciences} {\bfseries 1} (2017) 229}.

\bibitem{Krishnamoorthy2018Jul}
S.~Krishnamoorthy, E.~Blankemeyer, P.~Mollet, S.~Surti, R.~Van~Holen and
  J.S.~Karp, \emph{{Performance evaluation of the MOLECUBES
  {$\beta$}-CUBE{\ifmmode---\else\textemdash\fi}a high spatial resolution and
  high sensitivity small animal PET scanner utilizing monolithic LYSO
  scintillation detectors}},
  \href{https://doi.org/10.1088/1361-6560/aacec3}{\emph{Physics in Medicine
  {\&} Biology} {\bfseries 63} (2018) 155013}.

\bibitem{Yvon2020Jul}
D.~Yvon, V.~Sharyy, M.~Follin, J.-P.~Bard, D.~Breton, J.~Maalmi et~al.,
  \emph{{Design study of a scintronic crystal targeting tens of picoseconds
  time resolution for gamma ray imaging: the ClearMind detector}},
  \href{https://doi.org/10.1088/1748-0221/15/07/P07029}{\emph{J. Instrum.}
  {\bfseries 15} (2020) P07029}.

\bibitem{Stockhoff2021Jul}
M.~Stockhoff, M.~Decuyper, R.~Van~Holen and S.~Vandenberghe,
  \emph{{High-resolution monolithic LYSO detector with 6-layer
  depth-of-interaction for clinical PET}},
  \href{https://doi.org/10.1088/1361-6560/ac1459}{\emph{Physics in Medicine
  {\&} Biology} {\bfseries 66} (2021) 155014}.

\bibitem{Yvon2014}
D.~Yvon, J.-P.~Renault, G.~Tauzin, P.~Verrecchia, C.~Flouzat, S.~Sharyy et~al.,
  \emph{{CaLIPSO: An Novel Detector Concept for PET Imaging.}},
  \href{https://doi.org/10.1109/TNS.2013.2291971}{\emph{{IEEE} Transactions on
  Nuclear Science} {\bfseries 61} (2014) 60 }.

\bibitem{Hubbell2010XCOMP}
M.~Berger, J.~Hubbell, S.~Seltzer, J.~Chang, J.~Coursey, R.~Sukumar et~al.,
  ``Xcom : Photon cross sections database (version 1.5).''
  \url{http://physics.nist.gov/xcom}, (accessed May 17, 2024).

\bibitem{Ramos2015}
E.~Ramos, D.~Yvon, P.~Verrecchia, G.~Tauzin, D.~Desforge, V.~Reithinger et~al.,
  \emph{{Trimethyl Bismuth Optical Properties for Particle Detection and the
  CaLIPSO Detector}},
  \href{https://doi.org/10.1109/TNS.2015.2424080}{\emph{{IEEE} Transactions on
  Nuclear Science} {\bfseries 62} (2015) 1326}.

\bibitem{Ramos_2016}
E.~Ramos, O.~Kochebina, D.~Yvon, P.~Verrecchia, V.~Sharyy, G.~Tauzin et~al.,
  \emph{{Efficient and fast 511-{keV} $\gamma$ detection through Cherenkov
  radiation: the {CaLIPSO} optical detector}},
  \href{https://doi.org/10.1088/1748-0221/11/11/p11008}{\emph{J. Instrum.}
  {\bfseries 11} (2016) P11008}.

\bibitem{Kochebina2018Nov}
O.~Kochebina, S.~Jan, S.~Stute, V.~Sharyy, P.~Verrecchia, X.~Mancardi et~al.,
  \emph{{Performance Estimation for the High Resolution CaLIPSO Brain PET
  Scanner: A Simulation Study}},
  \href{https://doi.org/10.1109/TRPMS.2018.2880811}{\emph{IEEE Transactions on
  Radiation and Plasma Medical Sciences} {\bfseries 3} (2018) 363}.

\bibitem{Farradeche2019}
M.~Farradeche, G.~Tauzin, J.-P.~Mols, J.-P.~Bard, J.-P.~Dognon, C.~Weinheimer
  et~al., \emph{{Ionization parameters of Trimethylbismuth for high-energy
  photon detection}},
  \href{https://doi.org/10.1016/j.nima.2019.162646}{\emph{Nucl. Instrum. Meth.
  A} (2019) 162646}.

\bibitem{Gerke2022Sep}
B.~Gerke, S.-N.~Peters, N.~Marquardt, C.~Huhmann, V.M.~Hannen, M.~Holtkamp
  et~al., \emph{{Suppression of electrical breakdown phenomena in liquid
  TriMethyl Bismuth based ionization detectors}},
  \href{https://doi.org/10.1088/1748-0221/17/09/P09029}{\emph{Journal of
  Instrumentation} {\bfseries 17} (2022) P09029}.

\bibitem{Peters2022Sep}
S.~Peters, B.~Gerke, V.~Hannen, C.~Huhmann, N.~Marquardt,
  K.~Sch{\ifmmode\ddot{a}\else\"{a}\fi}fers et~al., \emph{{Electro-purification
  studies and first measurement of relative permittivity of TMBi}},
  \href{https://doi.org/10.48550/arXiv.2209.00996}{\emph{ArXiv e-prints} (2022)
  } [\href{https://arxiv.org/abs/2209.00996}{{\ttfamily 2209.00996}}].

\bibitem{SMARTGEL}
{Nye Lubricants}, ``{Technical Data Sheet for OCF-452}.''
  \url{https://www.nyelubricants.com/datasheet/TDS_SHORT_English_SMARTGEL+OCF-452+%28OBSOLETE%29.pdf},
  (accessed May 17, 2024).

\bibitem{Interconnect}
{Shin-Etsu Polymer}, ``{Inter-Connector\textsuperscript{\textregistered}
  MT-type}.''
  \url{https://www.shinetsu.info/product/mt-type-of-inter-connector}, (accessed
  May 17, 2024).

\bibitem{Delagnes2014Nov}
E.~Delagnes, D.~Breton, H.~Grabas, J.~Maalmi, P.~Rusquart and M.~Saimpert,
  \emph{{The SAMPIC Waveform and Time to Digital Converter}},  in \emph{{2014
  IEEE Nuclear Science Symposium and Medical Imaging Conference (NSS/MIC)}},
  pp.~1--9, IEEE, 2014,
  \href{https://doi.org/10.1109/NSSMIC.2014.7431231}{DOI}.

\bibitem{Delagnes:2015oda}
E.~Delagnes, D.~Breton, H.~Grabas, J.~Maalmi and P.~Rusquart, \emph{{Reaching a
  few picosecond timing precision with the 16-channel digitizer and timestamper
  SAMPIC ASIC}}, \href{https://doi.org/10.1016/j.nima.2014.12.042}{\emph{Nucl.
  Instrum. Meth. A} {\bfseries A787} (2015) 245}.

\bibitem{Breton2020}
D.~Breton, C.~Cheikali, E.~Delagnes, J.~Maalmi, P.~Rusquart and P.~Vallerand,
  \emph{{Fast electronics for particle Time-Of-Flight measurement, with focus
  on the SAMPIC ASIC}},
  \href{https://doi.org/10.1393/ncc/i2020-20007-6}{\emph{Nuovo Cimento C}
  {\bfseries 43} (2020) 7}.

\bibitem{Breton2016}
D.~Breton, V.~De~Cacqueray, E.~Delagnes, H.~Grabas, J.~Maalmi, N.~Minafra
  et~al., \emph{{Measurements of timing resolution of ultra-fast silicon
  detectors with the {SAMPIC} waveform digitizer}},
  \href{https://doi.org/10.1016/j.nima.2016.08.019}{\emph{Nucl. Instrum. Meth.
  A} {\bfseries 835} (2016) 51}.

\bibitem{Follin2022Mar}
M.~Follin, R.~Chyzh, C.-H.~Sung, D.~Breton, J.~Maalmi, T.~Chaminade et~al.,
  \emph{{High resolution MCP-PMT readout using transmission lines}},
  \href{https://doi.org/10.1016/j.nima.2021.166092}{\emph{Nucl. Instrum. Meth.
  A} {\bfseries 1027} (2022) 166092}.

\bibitem{LYSO}
{Luxium Solution, France}, ``{LYSO crystals, Technical data}.''
  \url{https://www.luxiumsolutions.com/radiation-detection-scintillators/crystal-scintillators/lyso-scintillation-crystals},
  (accessed May 17, 2024).

\bibitem{SENSL_SiPM}
ONSEMI, ``Silicon photomultipliers (sipm), low-noise, blue-sensitive c-series
  sipm sensors data sheet.''
  \url{https://www.onsemi.com/products/sensors/photodetectors-sipm-spad/silicon-photomultipliers-sipm/c-series},
  (accessed May 17, 2024).

\bibitem{PDG2022}
{\scshape Particle Data Group} collaboration, \emph{{Review of Particle
  Physics}}, \href{https://doi.org/10.1093/ptep/ptac097}{\emph{PTEP} {\bfseries
  2022} (2022) 083C01}.

\bibitem{agostinelli2003}
{\scshape GEANT4} collaboration, \emph{{GEANT4: A Simulation toolkit}},
  \href{https://doi.org/10.1016/S0168-9002(03)01368-8}{\emph{Nucl. Instrum.
  Meth. A} {\bfseries 506} (2003) 250}.

\bibitem{Allison2006Feb}
J.~Allison, K.~Amako, J.~Apostolakis, H.~Araujo, P.A.~Dubois, M.~Asai et~al.,
  \emph{{Geant4 developments and applications}},
  \href{https://doi.org/10.1109/TNS.2006.869826}{\emph{IEEE Trans. Nucl. Sci.}
  {\bfseries 53} (2006) 270}.

\bibitem{Allison2016Nov}
J.~Allison, K.~Amako, J.~Apostolakis, P.~Arce, M.~Asai, T.~Aso et~al.,
  \emph{{Recent developments in Geant4}},
  \href{https://doi.org/10.1016/j.nima.2016.06.125}{\emph{Nucl. Instrum. Meth.
  A} {\bfseries 835} (2016) 186}.

\bibitem{Sung_2023}
C.-H.~Sung, L.~Cappellugola, M.~Follin, S.~Curtoni, M.~Dupont, C.~Morel et~al.,
  \emph{Detailed simulation for the clearmind prototype detection module and
  event reconstruction using artificial intelligence},
  \href{https://doi.org/10.1016/j.nima.2023.168357}{\emph{Nucl. Instrum. Meth.
  A} {\bfseries 1053} (2023) 168357}.

\bibitem{canot2018}
C.~Canot, \emph{Fast and efficient optical Cherenkov detector for PET (in
  french)}, Ph.D. thesis, Paris-Saclay University, 2018.
\newblock \url{http://www.theses.fr/2018SACLS190}.

\bibitem{Canot2019Dec}
C.~Canot, M.~Alokhina, P.~Abbon, J.P.~Bard, D.~Breton, E.~Delagnes et~al.,
  \emph{{Fast and efficient detection of 511 keV photons using Cherenkov light
  in PbF2 crystal, coupled to a MCP-PMT and SAMPIC digitization module}},
  \href{https://doi.org/10.1088/1748-0221/14/12/p12001}{\emph{J. Instrum.}
  {\bfseries 14} (2019) P12001}.

\end{thebibliography}\endgroup

\end{document}